\documentclass[onecolumn,superscriptaddress,amsmath,amssymb]{revtex4}
\usepackage{graphicx}% Include figure files
\usepackage{dcolumn}% Align table columns on decimal point
\usepackage{bm}% bold math
\usepackage{color}
\usepackage{mathrsfs}
\usepackage{ulem}
\usepackage[german]{babel}

\def\bd{
\begin{document}} \def\ed{\end{document}}
\def\bmp{\begin{minipage}} \def\emp{\end{minipage}}
\def\bcc{\begin{center}} \def\ecc{\end{center}}     \def\npg{\newpage}
\def\beq{\begin{equation}} \def\eeq{\end{equation}} \def\hph{\hphantom}
\def\be{\begin{equation}} \def\ee{\end{equation}} \def\r#1{$^{[#1]}$}
\def\n{\noindent} \def\ni{\noindent} \def\pa{\parindent}
\def\hs{\hskip} \def\vs{\vskip} \def\hf{\hfill} \def\ej{\vfill\eject}
\def\cl{\centerline} \def\ob{\obeylines}  \def\ls{\leftskip}
\def\underbar#1{$\setbox0=\hbox{#1} \dp0=1.5pt \mathsurround=0pt
   \underline{\box0}$}   \def\ub{\underbar}    \def\ul{\underline}
\def\f{\left} \def\g{\right} \def\e{{\rm e}} \def\o{\over} \def\d{{\rm d}}
\def\vf{\varphi} \def\pl{\partial} \def\cov{{\rm cov}} \def\ch{{\rm ch}}
\def\la{\langle} \def\ra{\rangle} \def\EE{e$^+$e$^-$} \def\pt{p_{\rm t}}
\def\pti{p_{{\rm t},i}} \def\vti{v_{{\rm t},i}}
\def\ptj{p_{{\rm t},j}}\def\Pt{P_{\rm t}} \def\vt{v_{\rm t}}

\def\bitz{\begin{itemize}} \def\eitz{\end{itemize}}
\def\btbl{\begin{tabular}} \def\etbl{\end{tabular}}
\def\btbb{\begin{tabbing}} \def\etbb{\end{tabbing}}
\def\beqar{\begin{eqnarray}} \def\eeqar{\end{eqnarray}}
\def\\{\hfill\break} \def\dit{\item{-}} \def\i{\item}
\def\bbb{\begin{thebibliography}{9} \itemsep=-1mm}
\def\ebb{\end{thebibliography}} \def\bb{\bibitem}
\def\bpic{\begin{picture}(260,240)} \def\epic{\end{picture}}
\def\akgt{\cl{\bf ACKNOWLEDGMENTS}}
\def\fgn{\noindent{\bf\large\bf figure captions}}
%%%%%%%%%%%%%%%%%%%%%%%%%%%%%%%%%%%%%%%%%%%%%%%%%%%%%%%%%%%%%%%%%%%%%%
\def\m1{\langle N_p\rangle} \def\u2{\langle N_{\bar p}\rangle} \def\Nap{N_{\bar
p}}
%%%%%%%%%%%%%%%%%%%%%%%%%%%%%%%%%%%%%%%%%%%%%%%%%%%%%%%%%%%%%%%%%%%%%%%%%%%%%%
\def\lan{\langle}
\def\ran{\rangle}
\def\p{\pi}
\def\ifmath#1{\relax\ifmmode #1\else $#1$\fi}%
\def\rc{\ifmath{{\mathrm{c}}}}
\def\cut{\ifmath{{\mathrm{cut}}}}
\def\rF{\ifmath{{\mathrm{F}}}}
\def\rK{\ifmath{{\mathrm{K}}}}
\def\rp{\ifmath{{\mathrm{p}}}}
\def\rt{\ifmath{{\mathrm{t}}}}
\def\LAB{\ifmath{{\mathrm{LAB}}}}
\def\cut{\ifmath{{\mathrm{cut}}}}
\def\beq{\begin{equation}}
\def\eeq{\end{equation}}

\newcommand{\cinst}[2]{$^{\mathrm{#1}}$~#2\par}
\newcommand{\crefi}[1]{$^{\mathrm{#1}}$}
\newcommand{\crefii}[2]{$^{\mathrm{#1,#2}}$}
\newcommand{\crefiii}[3]{$^{\mathrm{#1,#2,#3}}$}
\newcommand{\HRule}{\rule{0.5\linewidth}{0.5mm}}

\bd
\title{Parameter dependence of the mapping of the mean-field Ising model onto QCD in the first-order phase transition regime}

\author{Xue Pan}\email{panxue1624@163.com}
\affiliation{School of Electronic Information and Electrical Engineering, Chengdu University, Chengdu 610106, China}
\author{Yuepeng Li}
\affiliation{School of Electronic Information and Electrical Engineering, Chengdu University, Chengdu 610106, China}

\begin{abstract}
%\sout{}
We map the mean-field Ising equation of state onto the QCD phase diagram and investigate how scaling parameters alter the metastable and unstable regions, as well as the symmetrized pressure and density along an isotherm on the QCD $T-\mu_B$ phase plane. We find that, for both the conventional mapping scheme and the scheme valid in the small-quark-mass limit, increasing the scaling parameters $w$ and $\rho$ narrows the metastable and unstable regions. In the small-quark-mass-limit scheme, larger $w$ and $\rho$ reduce both the Ising-symmetrized pressure and density. For the conventional scheme, increasing $w$ reduces these quantities, while increasing $\rho$ raises the Ising-symmetrized pressure and has only a small effect on the density.
\end{abstract}

\maketitle

\section{Introduction}

Constraining the scale of the quantum chromodynamics (QCD) phase diagram remains a central challenge in theoretical and experimental nuclear physics, with the search for the QCD critical point as a primary objective. In the region above the critical point, the transition from the hadronic phase to the quark–gluon plasma is a crossover~\cite{crossover1,crossover2}. At lower temperatures and higher baryon chemical potentials, a first-order phase transition is predicted, with the transition line terminating at the critical point~\cite{first1,first2,first3}.

In recent years, theoretical studies have predicted non‑monotonic fluctuations of conserved charges near the QCD critical point~\cite{Stephanov-prl102, Asakawa-prl103, Fuweijie, Vladi, Stephanov-prl107}. Significant fluctuations have also been measured in heavy-ion collision experiments~\cite{Phys.Rev.Lett.112.032302,Phys.Rev.Lett.135.142301}. Nevertheless, confirming the existence of the QCD critical point and pinning down its location remains a major challenge.

A family of equations of state for QCD are constructed with a critical point belonging to the three-dimensional Ising universality class~\cite{Stephanov-prc101,Anxin-NPA1017,Karthein-EPJP136}. The mapping from the Ising variables to the QCD coordinates relies on flexible parameters, such as the location and strength of the critical point. These thermodynamic properties lay the physical foundation for simulating fireball evolution in low-energy heavy-ion collisions and serve as essential inputs for hydrodynamic simulations. Meanwhile, conventional hydrodynamic frameworks need further refinement to properly characterize the propagation of critical fluctuations during QCD phase transitions.

To provide the necessary thermodynamic input for the first-order phase-transition region, the mean-field Ising model is mapped to QCD and used to study the coexistence region of the phase diagram, with its metastable and spinodal branches in Ref.~\cite{Karthein-PRD111}.
The conventional mapping scheme is adopted, in which the Ising temperature ($t$) axis is tangential to the QCD first-order phase transition line at the critical point, and the external magnetic field ($h$) axis is orthogonal to the 
$t$ axis~\cite{Asakawa-PRC71, PRC92.034912}. In this scheme, the critical point position is varied while the scaling parameters are held fixed~\cite{Karthein-PRD111}.

In fact, the $h$ axis will deform when mapped onto the QCD $T-\mu_B$ phase plane and it is not clear how this deformation occurs. This issue has been addressed in Ref.~\cite{Stephanov-PRD100}, where it is pointed out that in the limit of small quark masses, the angle between the $t$ axis and $h$ axis vanishes in the QCD $T-\mu_B$ phase plane as $m_q^{2/5}$, where $m_q\equiv m_u \approx m_d$ is the light quark mass. Moreover, in the random matrix model, it is found that the angles follow the relation $0<\alpha_2<\alpha_1$~\cite{Stephanov-PRD100}. Here $\alpha_1$ and $\alpha_2$ represent the angles between the horizontal axis (constant $T$) with the $t$ axis and $h$ axis, respectively, after they are mapped onto the QCD $T-\mu_B$ phase plane.

\begin{figure}[hbt]
	\centering
	\includegraphics[width=0.45\textwidth]{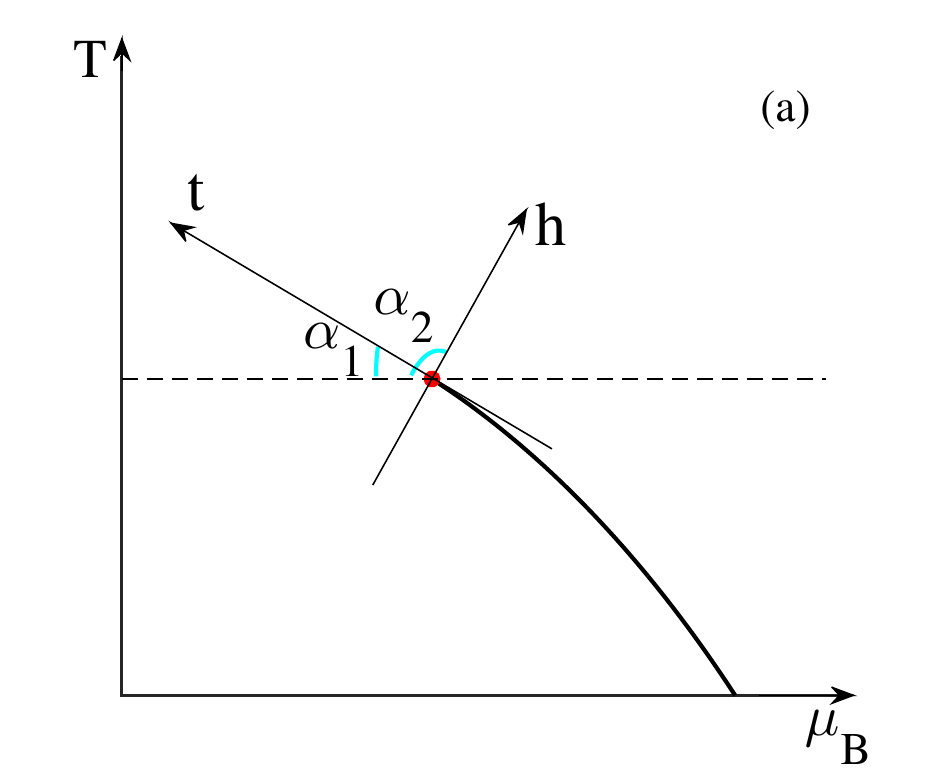}
	\includegraphics[width=0.45\textwidth]{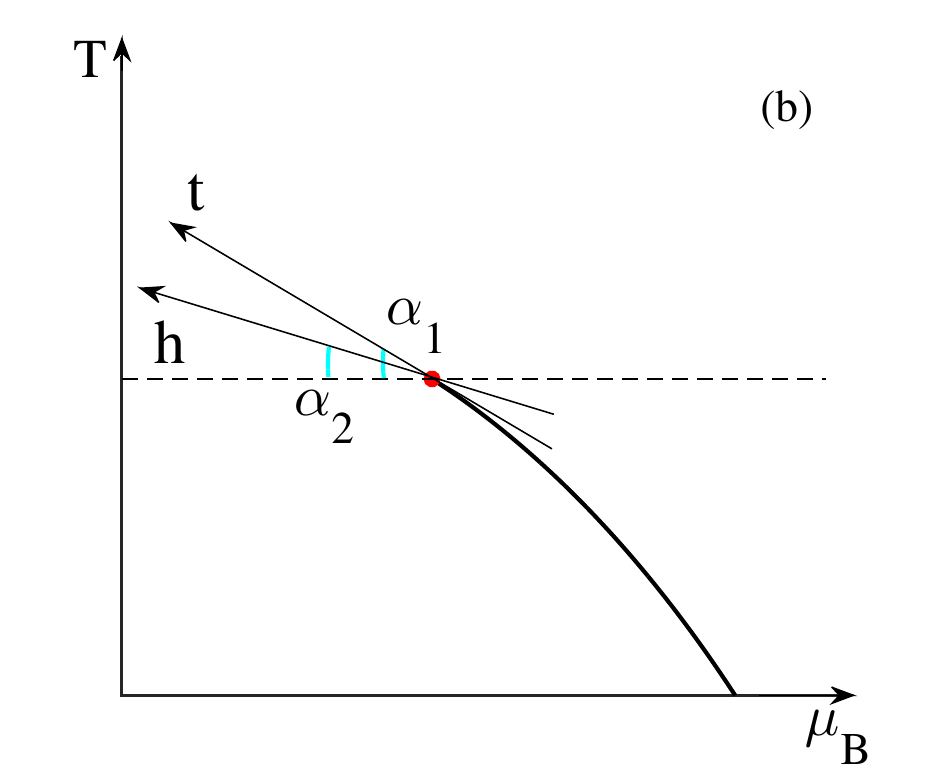}
	\caption{\label{Fig.1}(Color online). Sketches illustrating the mapping of Ising variables, reduced temperature $t$ and magnetic field $h$ onto the QCD $T-\mu_B$ phase plane in the common assumption (a) and in the small quark-mass limit (b). The solid red point represents the QCD critical point at the terminus of the first-order phase transition line. The black dashed line indicates the horizontal line at the critical temperature.}
\end{figure}

Sketches for the two different mapping schemes are shown in Fig.~1. The solid black line is the QCD first-order phase transition line and the solid red point represents the QCD critical point. In Fig.~1(a), the $h$ axis is orthogonal to the $t$ axis, that is $\alpha_2 = \alpha_1 + 90^{\circ}$. In Fig.~1(b), $0<\alpha_2<\alpha_1$. 

Different mapping schemes affect the behavior of thermodynamic quantities. As shown in Refs.~\cite{Stephanov-PRC103,panxue-CPC49,panxue-IJMPE34}, fluctuations of conserved charges display different behaviors for different mapping schemes. Within the same mapping scheme, the fluctuations are sensitive to the scaling parameters.

These parameters are also expected to influence the global structure of the phase diagram. So it is necessary to investigate, within the mean-field Ising model, how different mapping schemes and scaling parameters affect the phase diagram structure and the thermodynamic evolution.

In this paper, following recent theoretical predictions~\cite{chiraltransitionT1,chiraltransitionT2,Fuweijie-PRD101,Fischer,Isserstedt,Gaofei1,Gaofei2}, we fix the net-baryon chemical potential of the critical point ($\mu_{BC}$) at $630$ MeV, and study the influence of different mapping schemes and scaling parameters: how these parameters reshape the geometry of coexistence lines, modify the extent of metastable and spinodal unstable regions, and alter the characteristic thermodynamic quantities.

The paper is organized as follows. In Section 2, we introduce the mean-field Ising model and discuss the linear mapping from the mean-field Ising model to QCD. In Section 3, we investigate how different mapping schemes and scaling parameters affect the metastable and unstable regions, as well as the pressure and density. Finally, Section 4 presents the conclusions.

\section{Linear mapping from the mean-field Ising model to QCD}

In the mean-field Ising model, the Helmholtz free energy $F$, external magnetic field $h$, and pressure $P$, all expressed as functions of the magnetization $M$ and reduced temperature $t$, are defined as follows,
\begin{equation}\label{Helmholtz free energy}
	F(M,t)=h_0m_0(\frac{3}{2}t\tilde{M}^2+\frac{1}{4}\tilde{M}^4),
\end{equation}
\begin{equation}\label{external magnetic field}
	h(M,t)=\left(\frac{\partial{F}}{\partial{M}}\right)_t=h_0\tilde{h}(M,t)=h_0(3t\tilde{M}+\tilde{M}^3),
\end{equation}
\begin{equation}\label{pressure}
	P(M,t)=Mh-F(M,t)=h_0m_0(\frac{3}{2}t\tilde{M}^2+\frac{3}{4}\tilde{M}^4),
\end{equation}
where $\tilde{M}=M/m_0$, and $m_0$ and $h_0$ are normalization constants. These constants are fixed by imposing $M(r=-1,h=0+)=1$
and $M(r=0, h)\propto \mathrm{sgn}(h)|h|^{1/\delta}$, yielding $h_0=\frac{1}{3\sqrt{3}}$, $m_0=\frac{1}{\sqrt{3}}$~\cite{Asakawa-PRC71, Koch-RP853}.

As shown in Eq.~\eqref{external magnetic field}, $h$ is a cubic function defined in terms of $M$ and $t$. To uncover the properties of spinodal points, it is necessary to further study the relation between $M$ and $h$, and characterize the mean-field Ising equation of state via the behavior of $M(h)$. To this end, the equation $h(M,t)$ is inverted, yielding three solutions. We write $\tilde{h}(M,t)=\tilde{M}(\tilde{M}+i\sqrt{3t})(\tilde{M}-i\sqrt{3t})$. Then one real solution and an additional pair of complex solutions for $\tilde{M}(h,t)$ are yielded,
\begin{equation}\label{magnetization}
	\tilde{M}(h,t)=
	\left\{ \begin{array}{ll}
		\dfrac{2^{1/3}}{\eta^{1/3}}t-\dfrac{\eta^{1/3}}{2^{1/3}}
		\\  -\dfrac{(1+i\sqrt{3})t}{2^{2/3}\eta^{1/3}}+\dfrac{(1-i\sqrt{3})\eta^{1/3}}{2^{4/3}}
		\\-\dfrac{(1-i\sqrt{3})t}{2^{2/3}\eta^{1/3}}+\dfrac{(1+i\sqrt{3})\eta^{1/3}}{2^{4/3}},
	\end{array} \right. 
\end{equation}
where $\eta=\sqrt{\tilde{h}^{2}+4t^3}-\tilde{h}$. Both this inversion procedure and the resulting solutions can be found in Ref.~\cite{Karthein-PRD111}. Using Eq.~\eqref{magnetization}, it is convenient to express the thermodynamic quantity $P$ in terms of $h$ and $t$.

In the mean‑field Ising model, the coexistence points and spinodal points all lie below the critical temperature ($t=0$). At the coexistence points, the Helmholtz free energy exhibits minima satisfying $(\partial F/\partial M)_t=0$. It is evident that these coexistence points correspond to $h=0$. From the spinodal definition, the spinodal points can be obtained by solving the equation $(\partial h/\partial M)_t=0$, which gives 
\begin{equation}
	h_{\mathrm{sp}}=\pm \frac{2t^{3/2}}{3\sqrt{3}}.
\end{equation}

The phase diagram of mean-field Ising model including the critical point, coexistence line and the spinodal lines are shown in Fig.~2. 
The red point represents the critical point. The solid black line is the first-order coexistence line. The two dashed blue lines are the spinodal lines.

\begin{figure*}[hbt]
	\centering
	\includegraphics[width=0.45\textwidth]{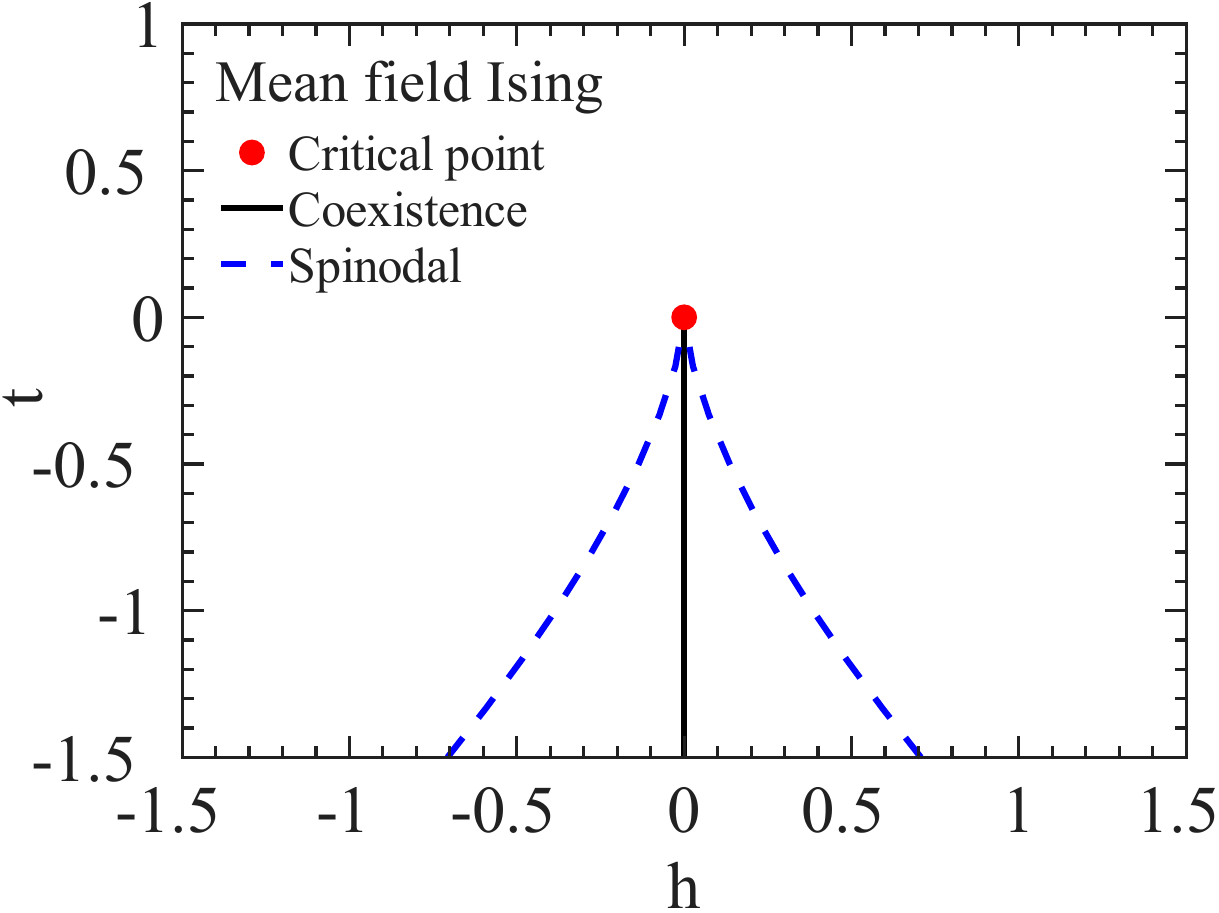}
	\caption{\label{Fig. 2}(Color online). Phase diagram of the mean-field Ising model, showing the critical point (red solid point), the coexistence line (solid black line), and the spinodal lines (dashed blue lines).}
\end{figure*}

To map the mean-field Ising resluts onto QCD, a linear mapping is employed as in Ref.~\cite{Stephanov-prc101},
\begin{equation}\label{linear map}
	\frac{T-T_C}{T_C}=w(\rho t\sin\alpha_1+h\sin\alpha_2),
\end{equation}
\begin{equation}\label{linear map1}
	\frac{\mu_B-\mu_{BC}}{T_C}=w(-\rho t\cos\alpha_1-h\cos\alpha_2),
\end{equation}
where $T_C$ and $\mu_{BC}$ are the temperature and net-baryon chemical potential at the QCD critical point, respectively. $w$ and $\rho$ are two scaling parameters of the mapping from Ising model to QCD. $\alpha_1$ and $\alpha_2$ are two angles which have been introduced in the introduction, as shown in Fig.~1. 

To reduce the number of mapping parameters, we utilize results from the Wuppertal-Budapest collaboration on the chiral phase transition line up to $O(\mu_B^4)$~\cite{Fodor-PRL125}, 
\begin{equation}\label{QCD transition line}
	T= T_0[1-\kappa_2(\frac{\mu_B}{T_0})^2-\kappa_4(\frac{\mu_B}{T_0})^4],
\end{equation}
where $T_0$ is the transition temperature at $\mu_B=0$, which is set as $158$ MeV. $\kappa_2$ is the curvature at $\mu_B=0$, which is set as $0.0153$. $\kappa_4$ is set as $0.00032$. Assuming the critical point lies on the chiral phase transition line, the QCD critical temperature $T_C$ is $106.8$ MeV for a fixed critical net-baryon chemical potential $\mu_{BC}=630$ MeV.

Because the $t$ axis in the Ising model is tangential to the first-order phase transition line at the QCD critical point, the value of $\alpha_1$ can be fixed at $11.5^{\circ}$. In the conventional mapping scheme shown in Fig.~1(a), the $h$ axis is orthogonal to the $t$ axis. Then $\alpha_2=\alpha_1+90^{\circ}=101.5^{\circ}$. For the case illustrated in Fig.~1(b), we set $\alpha_2=\alpha_1-6^{\circ}=5.5^{\circ}$. 

After mapping the Ising variables to the QCD coordinates by Eqs.~\eqref{linear map} and \eqref{linear map1}, the spinodal points in the QCD $T-\mu_B$ phase diagram can be obtained,

\begin{equation}\label{musp}
	\begin{aligned}
		&\frac{-(\mu_{B,\mathrm{sp}}-\mu_{BC})\sin{\alpha_1}-(T-T_C)\cos{\alpha_1}}{T_Cw\sin(\alpha_1-\alpha_2)}= \\ &\pm\frac{2}{3\sqrt{3}}\left(\frac{(\mu_{B,\mathrm{sp}}-\mu_{BC})\sin{\alpha_2}+(T-T_C)\cos{\alpha_2}}{T_Cw\rho\sin(\alpha_1-\alpha_2)}\right)^{3/2}.
	\end{aligned}
\end{equation}

When $\alpha_2=\alpha_1+90^{\circ}$, we can reduce Eq.~\eqref{musp} to the following form,

\begin{equation}\label{musp1}
	\begin{aligned}
		&\frac{(\mu_{B,\mathrm{sp}}-\mu_{BC})\sin{\alpha_1}+(T-T_C)\cos{\alpha_1}}{T_Cw}= \\ &\pm\frac{2}{3\sqrt{3}}\left(\frac{-(\mu_{B,\mathrm{sp}}-\mu_{BC})\cos{\alpha_1}+(T-T_C)\sin{\alpha_1}}{T_Cw\rho}\right)^{3/2}.
	\end{aligned}
\end{equation}

\begin{figure*}[hbt]
	\centering
	\includegraphics[width=0.32\textwidth]{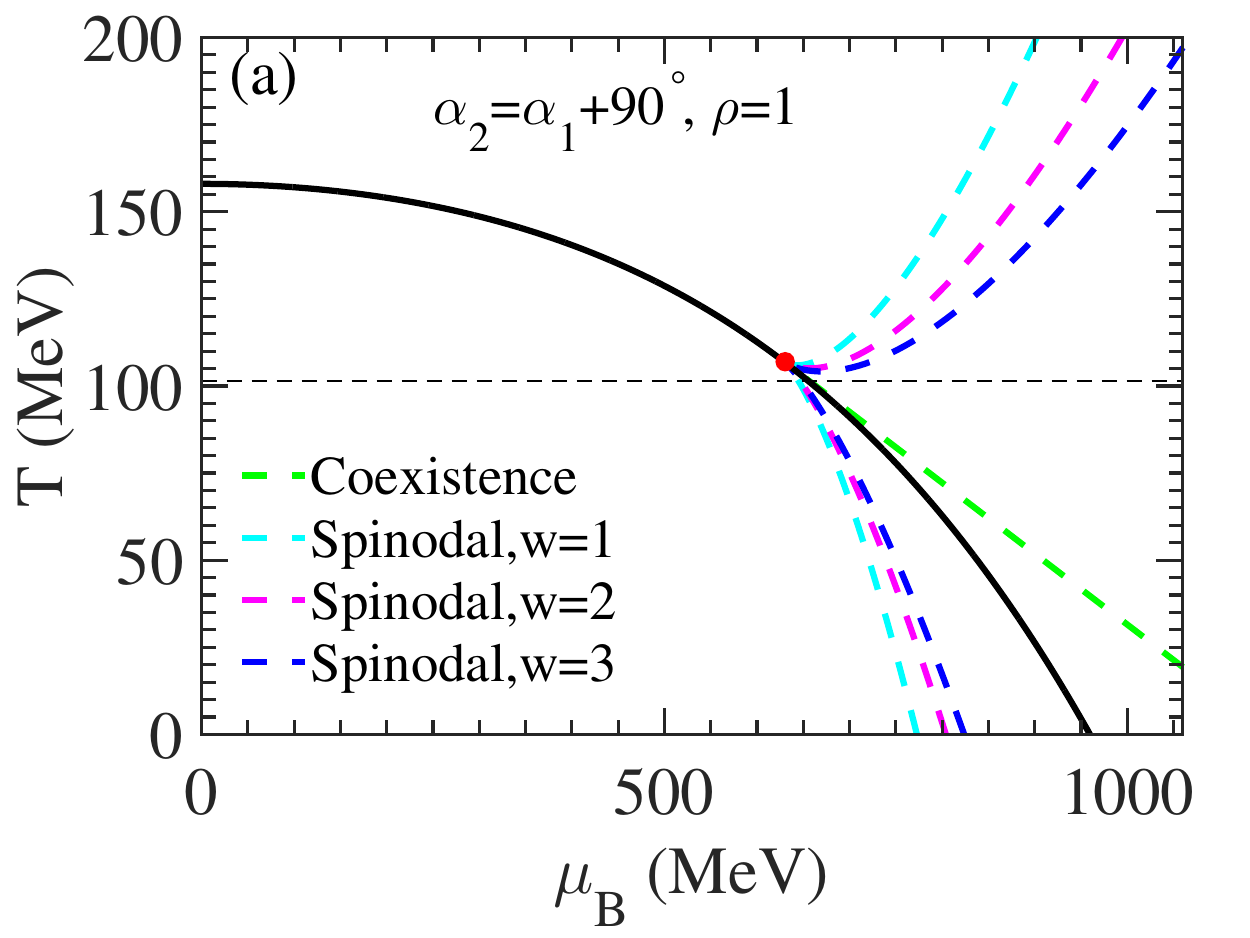}
	\includegraphics[width=0.32\textwidth]{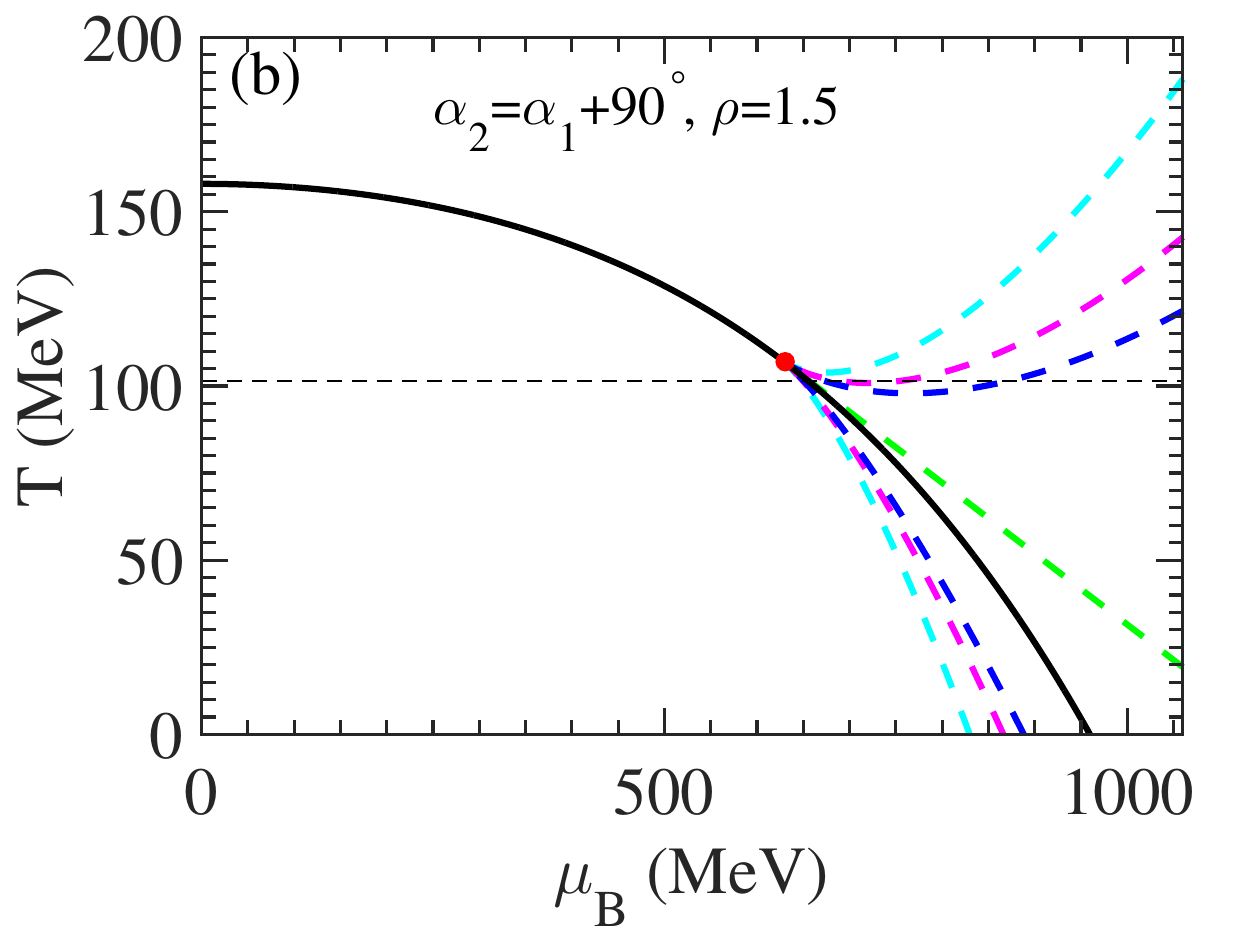}
	\includegraphics[width=0.32\textwidth]{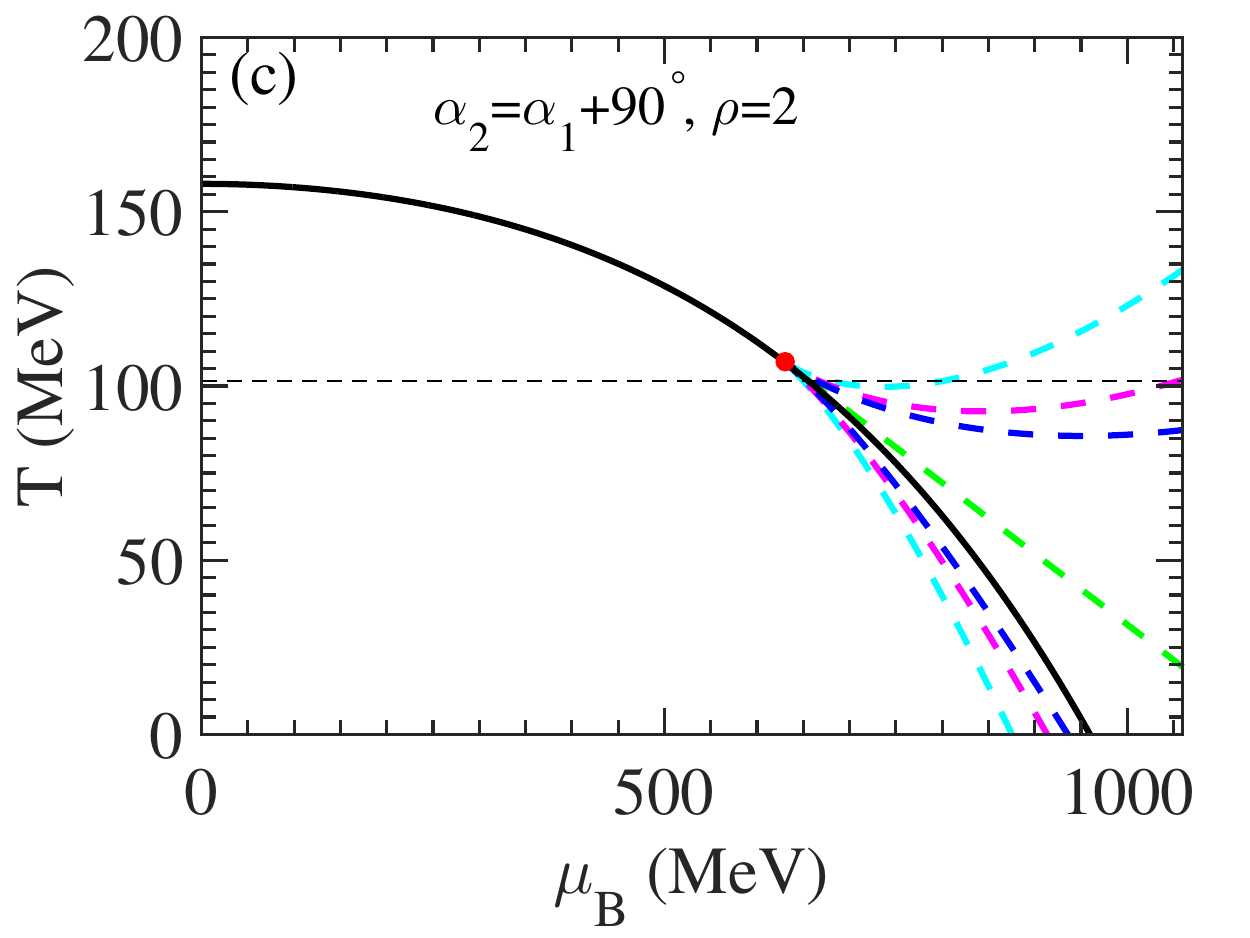}
	\caption{\label{Fig. 3}(Color online). Phase diagram of QCD on the $T-\mu_B$ phase plane, showing spinodal lines mapped from the mean-field Ising model for $\alpha_2=\alpha_1+90^\circ$ at different scaling parameters. The solid black line denotes the chiral phase transition line given in Eq.~\eqref{QCD transition line}. The red point indicates the QCD critical point. }
\end{figure*}

\begin{figure*}[hbt]
	\centering
	\includegraphics[width=0.32\textwidth]{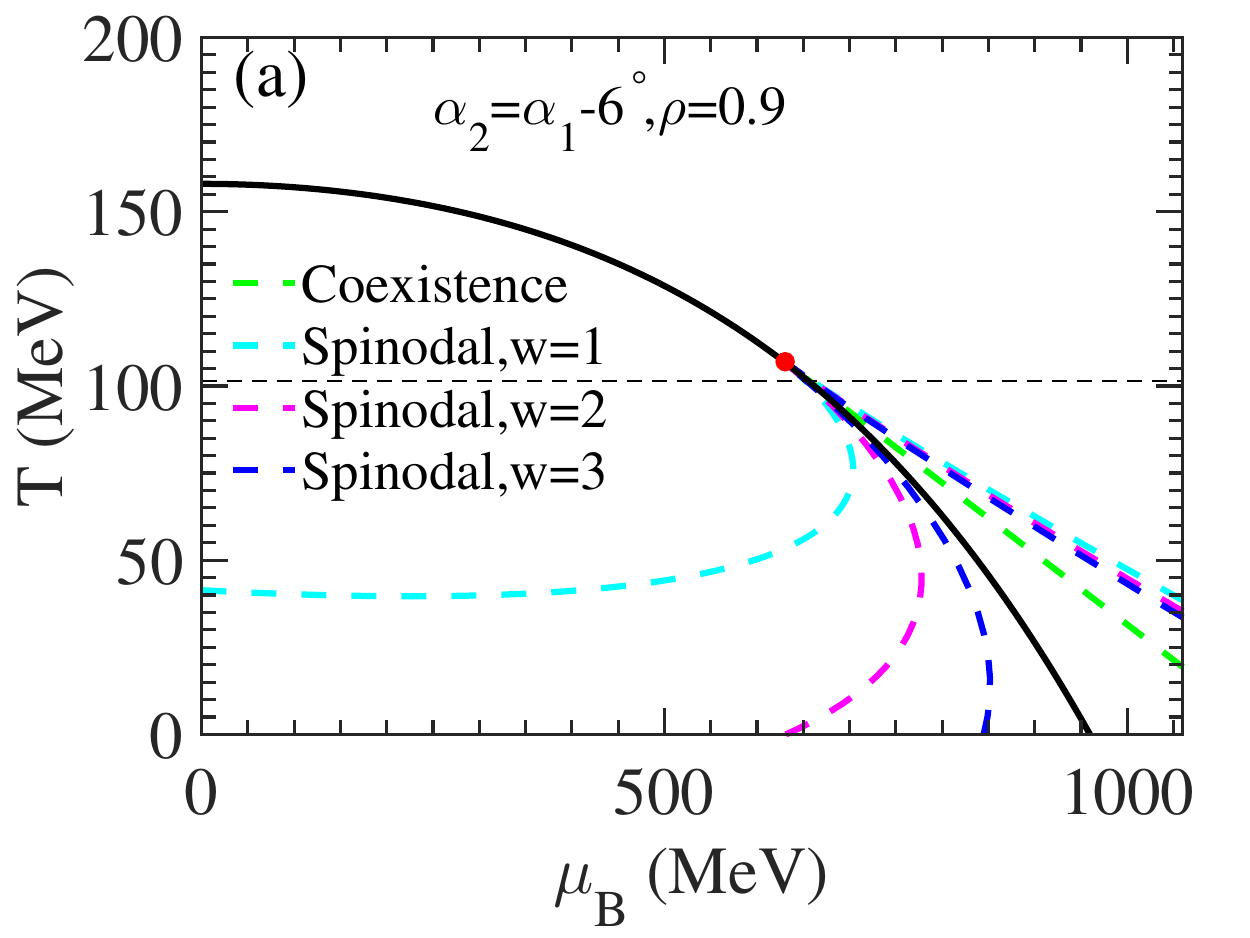}
	\includegraphics[width=0.32\textwidth]{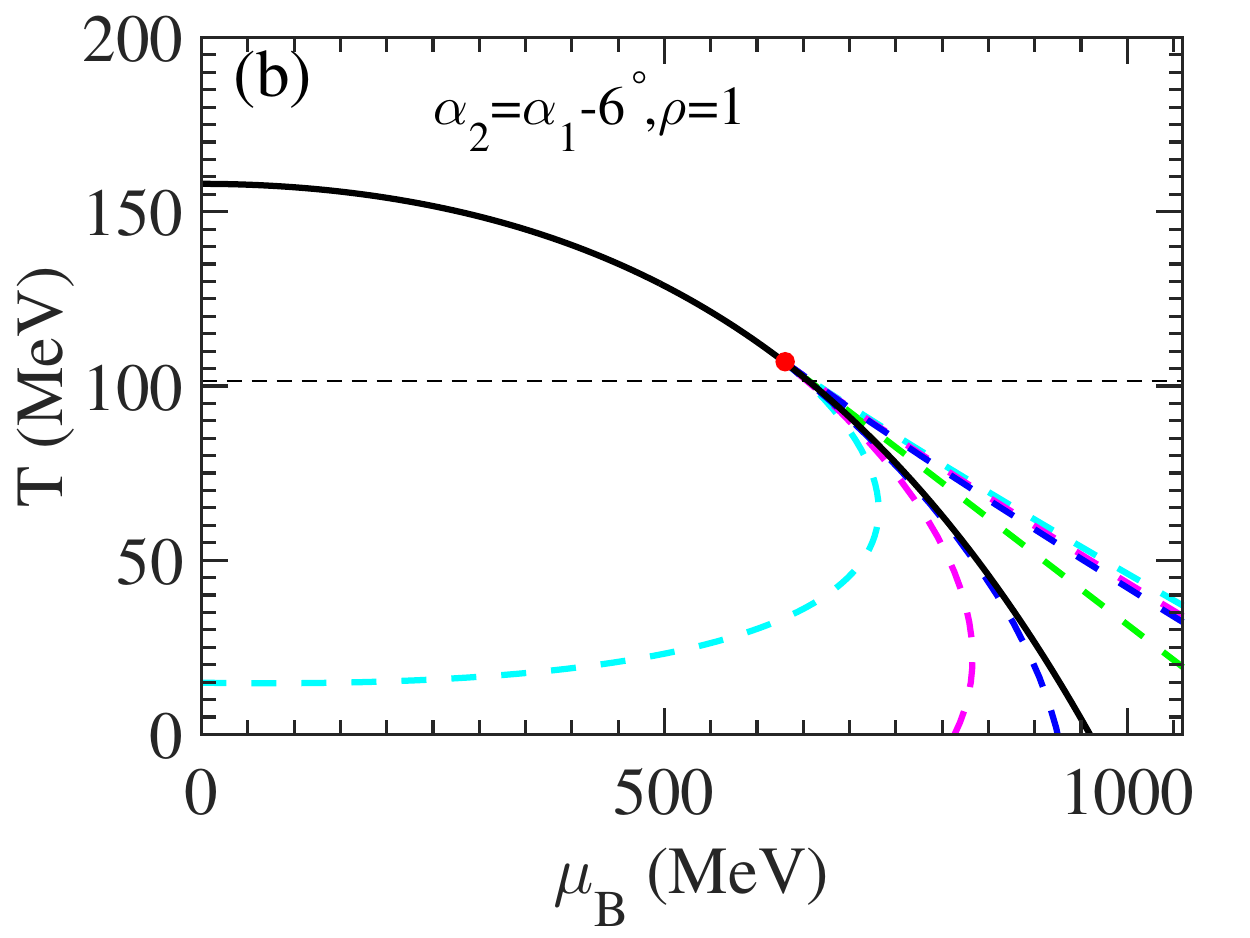}
	\includegraphics[width=0.32\textwidth]{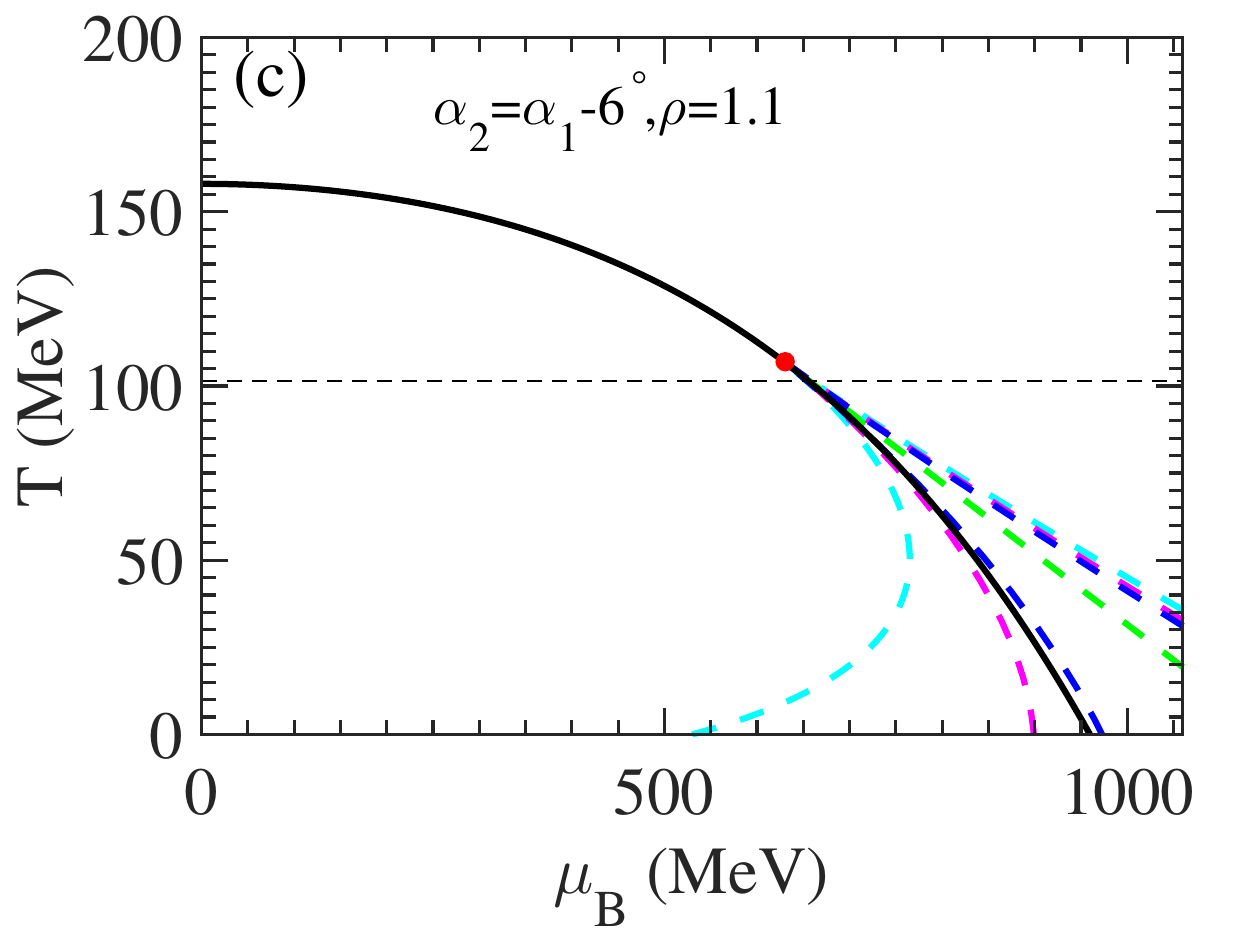}
	\caption{\label{Fig. 4}(Color online). Phase diagram of QCD on the $T-\mu_B$ phase plane, showing spinodal lines mapped from the mean-field Ising model for $\alpha_2=\alpha_1-6^\circ$ at several values of the scaling parameters. The solid black line denotes the chiral phase transition line given in Eq.~\eqref{QCD transition line}. The red point indicates the QCD critical point.}
\end{figure*}

\section{Effects of the mapping parameters}

For the first kind of mapping scheme as showed in Fig.~1(a), that is $\alpha_2=\alpha_1+90^\circ$, the influence of the scaling parameters of $w$ and $\rho$ on the spinodal lines are shown in Fig.~3. The black solid line represents the chiral phase transition line in Eq.~\eqref{QCD transition line}, the red point represents the QCD critical point at $(T_C, \mu_{BC})=(106.8, 630)$ MeV. The black horizontal dashed line represents the isotherm at $0.95T_{C}$ and serves as a guide to the eye.

The green dashed line presents the coexistence line mapped from the mean-field Ising model. Its position does not vary as the mapping parameters. 

In each subfigure of Fig.~3, $\rho$ is fixed, and the spinodal lines shift toward each other and toward the coexistence line as $w$ increases; i.e., the two blue dashed lines are closest to each other and also lie nearest to the coexistence line, while the two cyan dashed lines are the most distant to each other and also lie farthest from the coexistence line.

For fixed $w$, as $\rho$ increases, the spinodal lines also shift toward each other and toward the coexistence line. For example, the cyan dashed lines in Fig.~3(c) are closest to each other and the coexistence line, while those in Fig.~3(a) are farthest from each other and the coexistence line. The same trend holds for the purple and blue dashed lines. 

\begin{figure*}[hbt]
	\centering
	\includegraphics[width=0.32\textwidth]{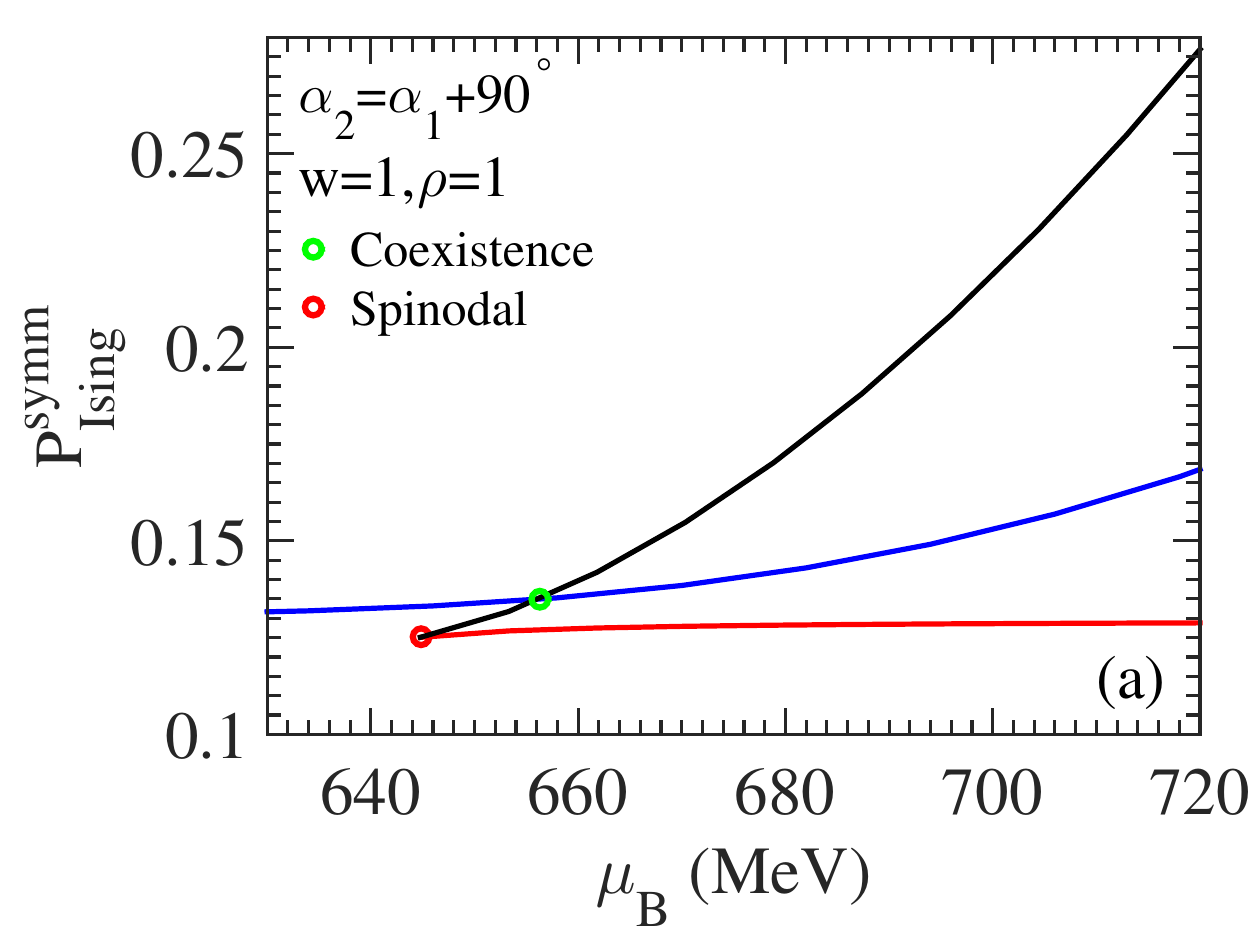}
	\includegraphics[width=0.32\textwidth]{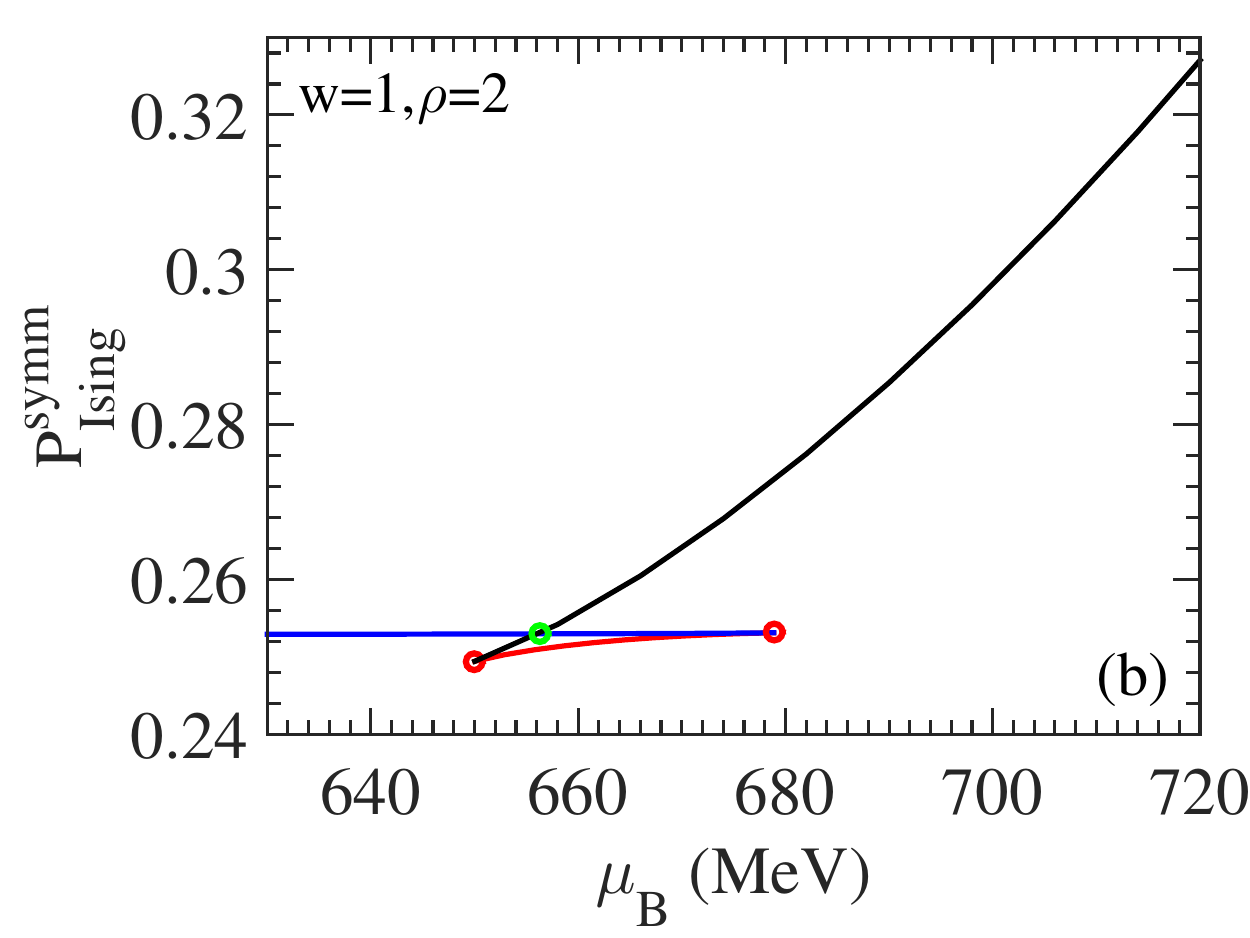}
	\includegraphics[width=0.32\textwidth]{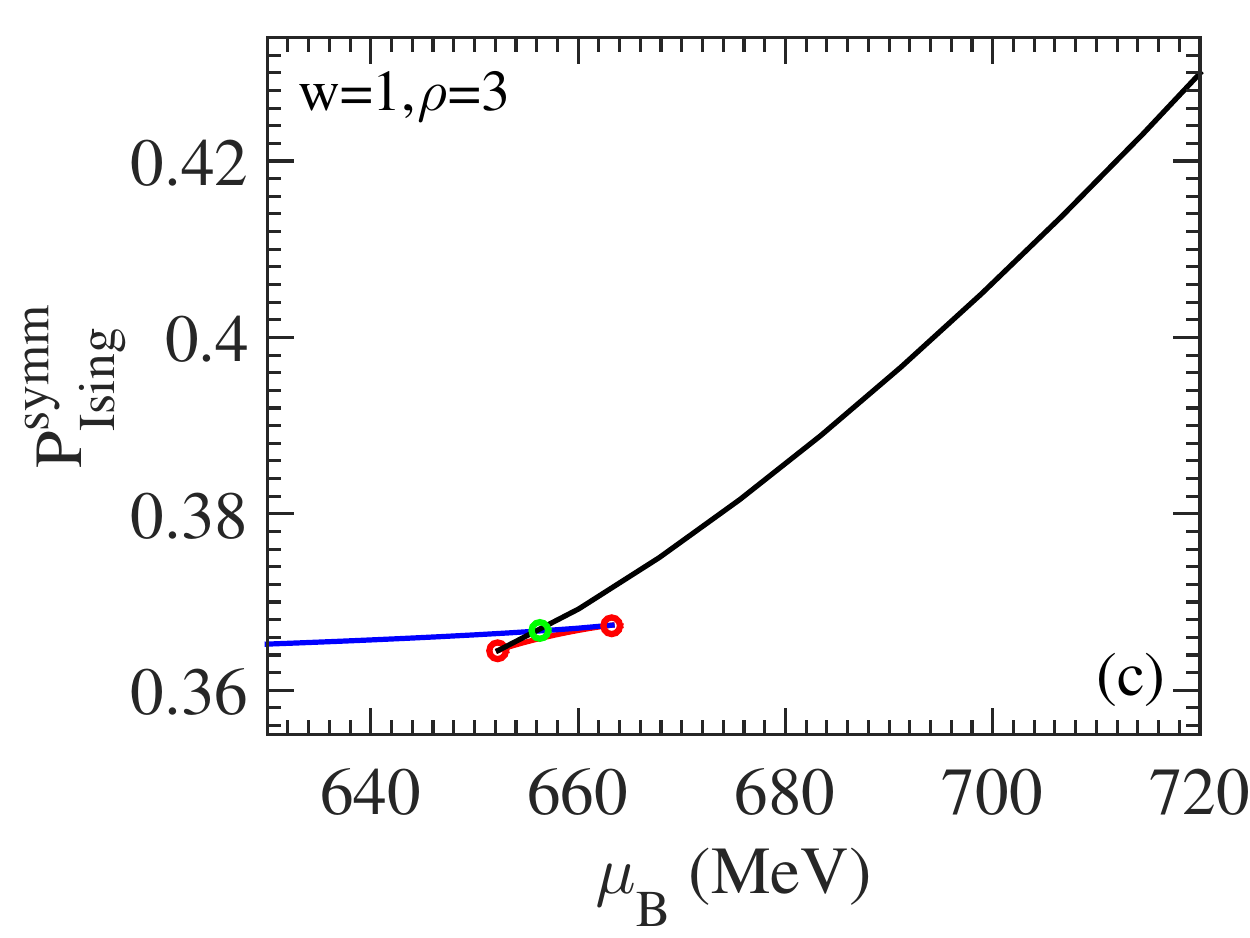}
	\includegraphics[width=0.32\textwidth]{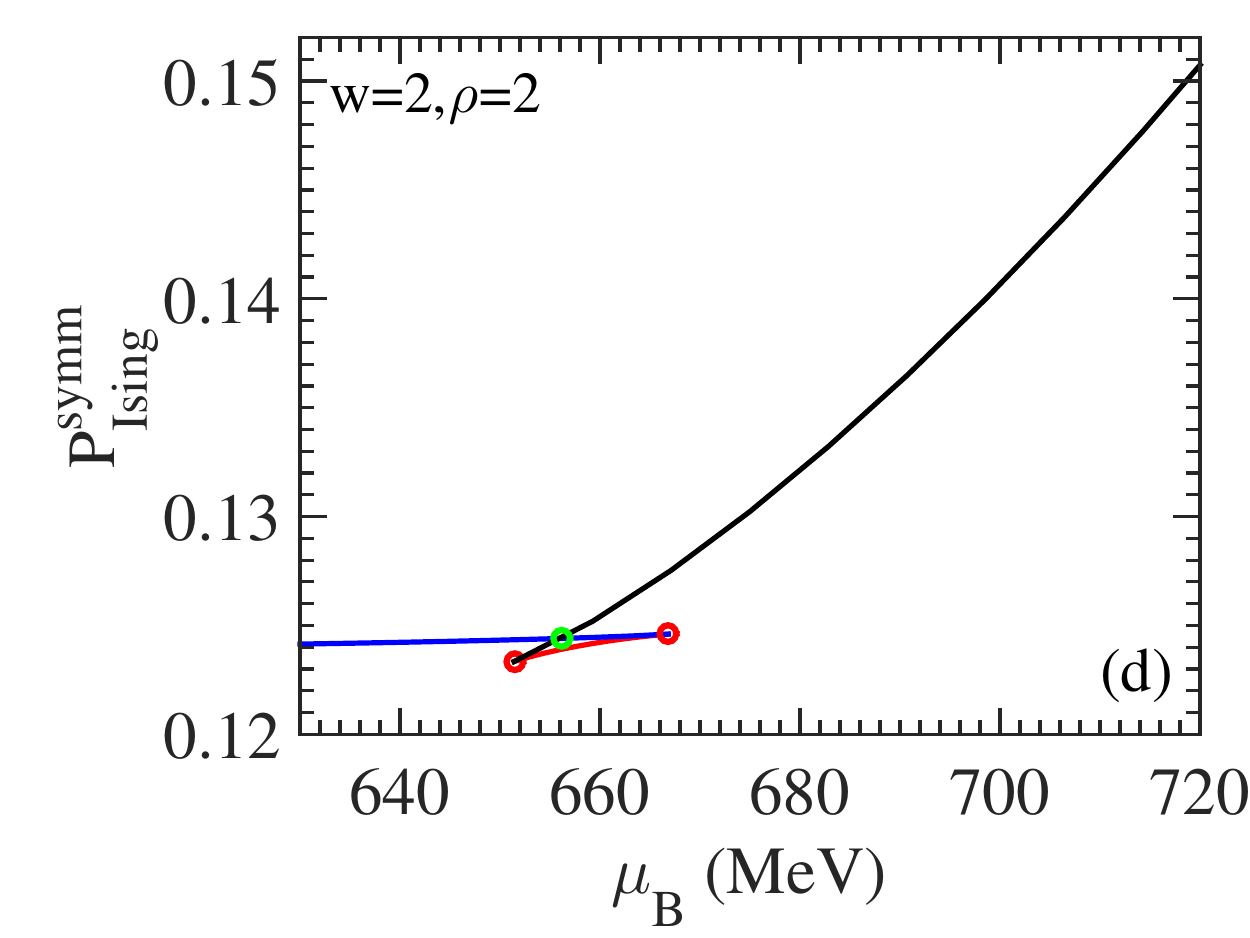}
	\includegraphics[width=0.32\textwidth]{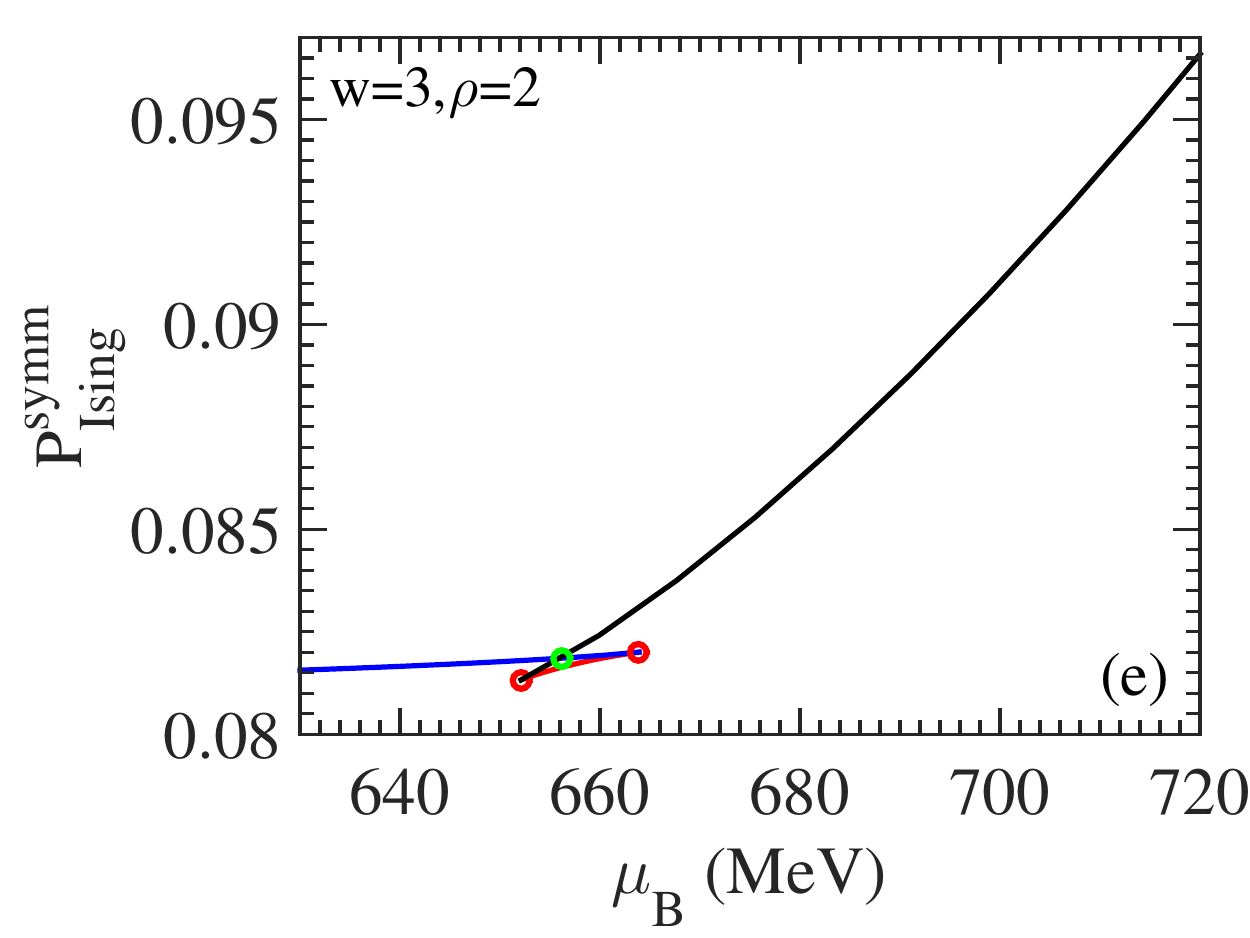}
	
	\caption{\label{Fig. 5}(Color online). Symmetrized Ising model pressure as a function of the net-baryon chemical potential for $\alpha_2=\alpha_1+90^\circ$ at different scaling parameters. The open green and red circles indicate the location of the coexistence and spinodal points, respectively.}
\end{figure*}

On the other hand, in Fig.~3(a), where $\rho$ is fixed at $1$ and $w$ varies from $1$ to $3$, with an intermediate value of $2$, only one intersection is observed between the isotherm and spinodal lines below the coexistence line, whereas no such intersection exists above it. When $\rho$ is raised to $1.5$ as shown in Fig.~3(b), one more intersection emerges above the coexistence lines for $w=2$ and $3$. Further increasing $\rho$ to $2$ in Fig.~3(c) leads to an intersection above the coexistence lines for all $w=1$, $2$ and $3$. 

Thus for small $w$ and $\rho$, only one spinodal point can be observed along the isotherm; when $w$ and $\rho$ are sufficiently large, two spinodal points appear.

For $\alpha_2=\alpha_1-6^{\circ}$, spinodal lines varying with the scaling parameters are shown in Fig.~4. The spinodal lines shift toward each other and toward the coexistence line as $w$ or $\rho$ increases. These trends are the same as those for $\alpha_2=\alpha_1+90^\circ$.

Comparing Fig.~4(b) with Fig.~3(a), one can see that for $w=1$ and $\rho=1$, the spinodal lines lie closer to each other and also the coexistence line for $\alpha_2=\alpha_1-6^\circ$ near the critical point. There are two intersections between the spinodal line with the isotherm at $0.95T_{C}$ for $\alpha_2=\alpha_1-6^{\circ}$, while only one for $\alpha_2=\alpha_1+90^\circ$. 

In each subfigure of Fig.~4, the spinodal lines above the coexistence line lie closer to the coexistence line than those below it, whereas the opposite trend is observed in each subfigure of Fig.~3.

Qualitatively, increasing either $w$ or $\rho$ causes the spinodal lines to approach each other and also the coexistence line for both mapping schemes shown in Fig.~1. The difference is that, for the same value of $w$ and $\rho$, the spindal lines lie closer to each other and the coexistence line in the vicinity of the critical point when $\alpha_2=\alpha_1-6^\circ$. 

Turning to thermodynamic properties, since QCD exhibits symmetry under the exchange of matter and anti-matter, charge-conjugation symmetry must hold across the $ \nu_B= 0$ line; for further details, please see Ref.~\cite{Karthein-PRD111}. The pressure and density are symmetrized as follows,
\begin{equation}\label{Psymm}
	P_{\mathrm{symm}}(T,\mu_B)=\frac{1}{2}(P(T,\mu_B)+P(T,-\mu_B)),
\end{equation}
\begin{equation}\label{nBsymm}
	n_{B,\mathrm{symm}}(T,\mu_B)=\frac{1}{2}(n_B(T,\mu_B)-n_B(T,-\mu_B)),
\end{equation}
where $n_B(T,\mu_B)=\left.\frac{\partial P(T,\mu_B)}{\partial \mu_B}\right|_{T}$.

The $\mu_B$ dependence of symmetrized pressure from the Ising model ($P_{\mathrm{Ising}}^{\mathrm{symm}}$) after mapping to the QCD phase diagram with $\alpha_2=\alpha_1+90^\circ$ is shown in Fig.~5. The $\mu_B$ values range from $630$ to $720$ MeV. The temperature is fixed at $0.95T_C$.

The three pressure solutions, derived from the three‑branch solutions for $M(h)$ in Eq.~\eqref{magnetization}, are plotted in distinct colors. The coexistence region starts where the first two solutions (plotted in blue and black) intersect for $\mu_B > \mu_{BC}$, and the coexistence point is marked by the open green circle. The spinodal points occur where these two solutions meet the third solution (plotted in red) and are marked by the open red circle.

\begin{figure*}[hbt]
	\centering
	\includegraphics[width=0.32\textwidth]{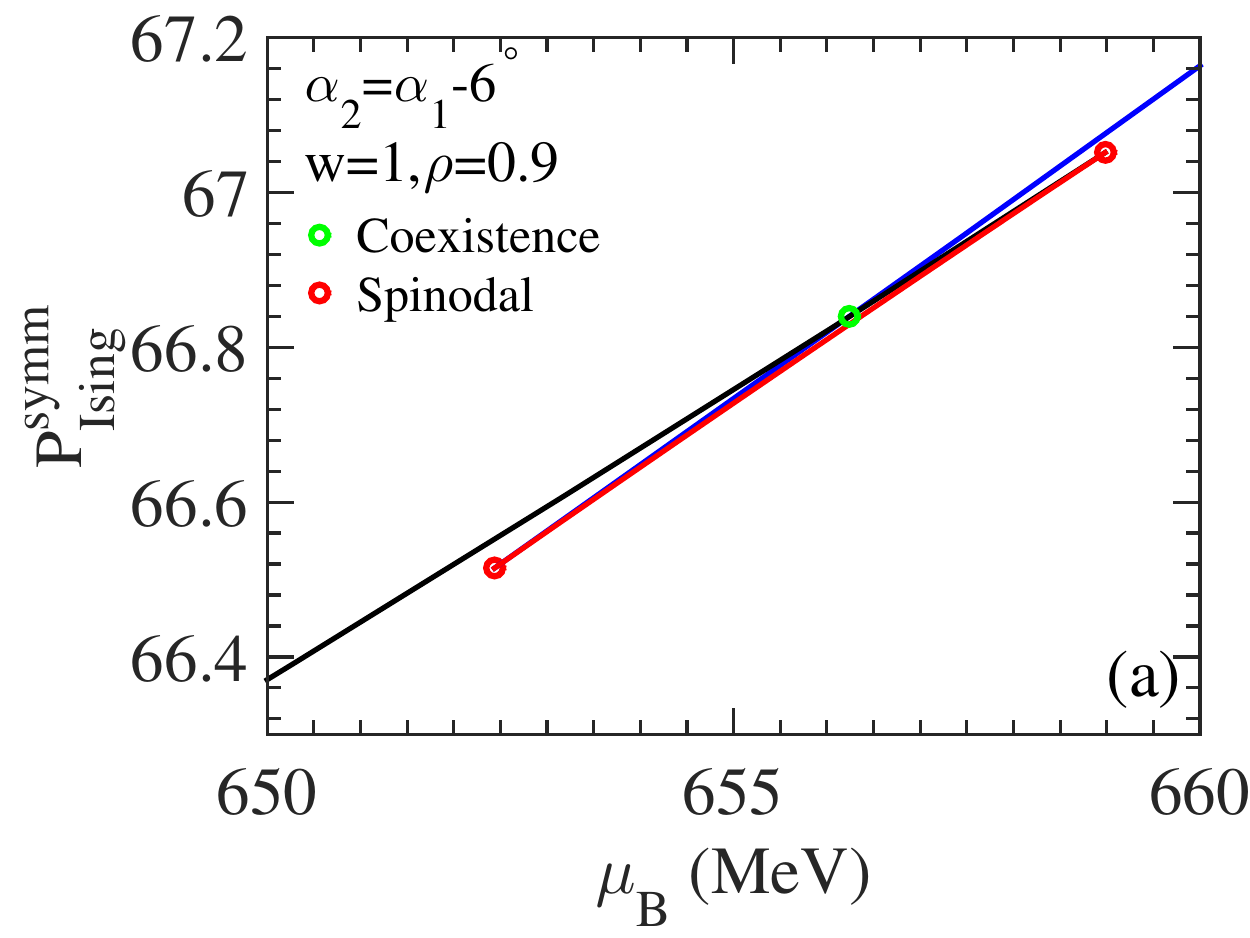}
	\includegraphics[width=0.32\textwidth]{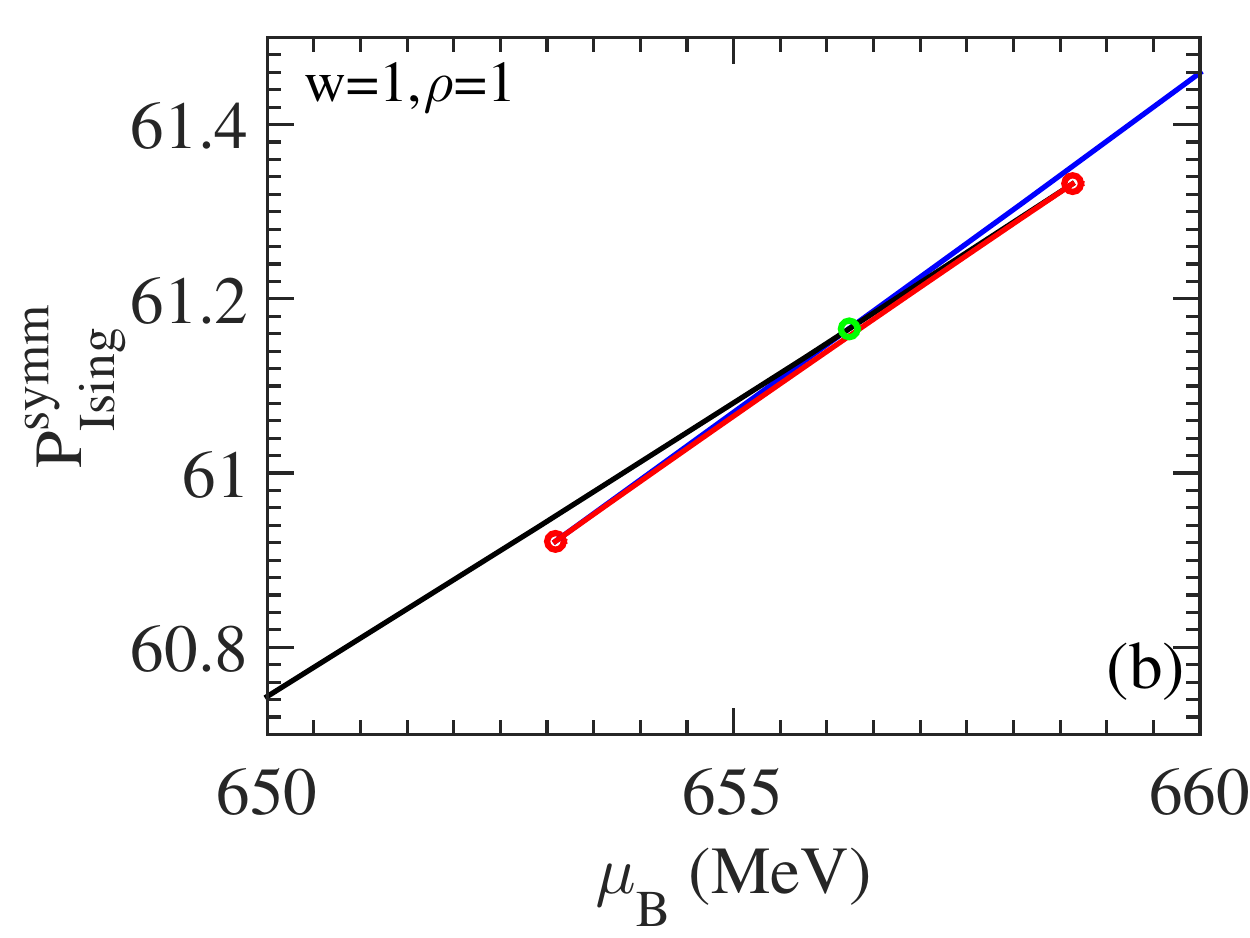}
	\includegraphics[width=0.32\textwidth]{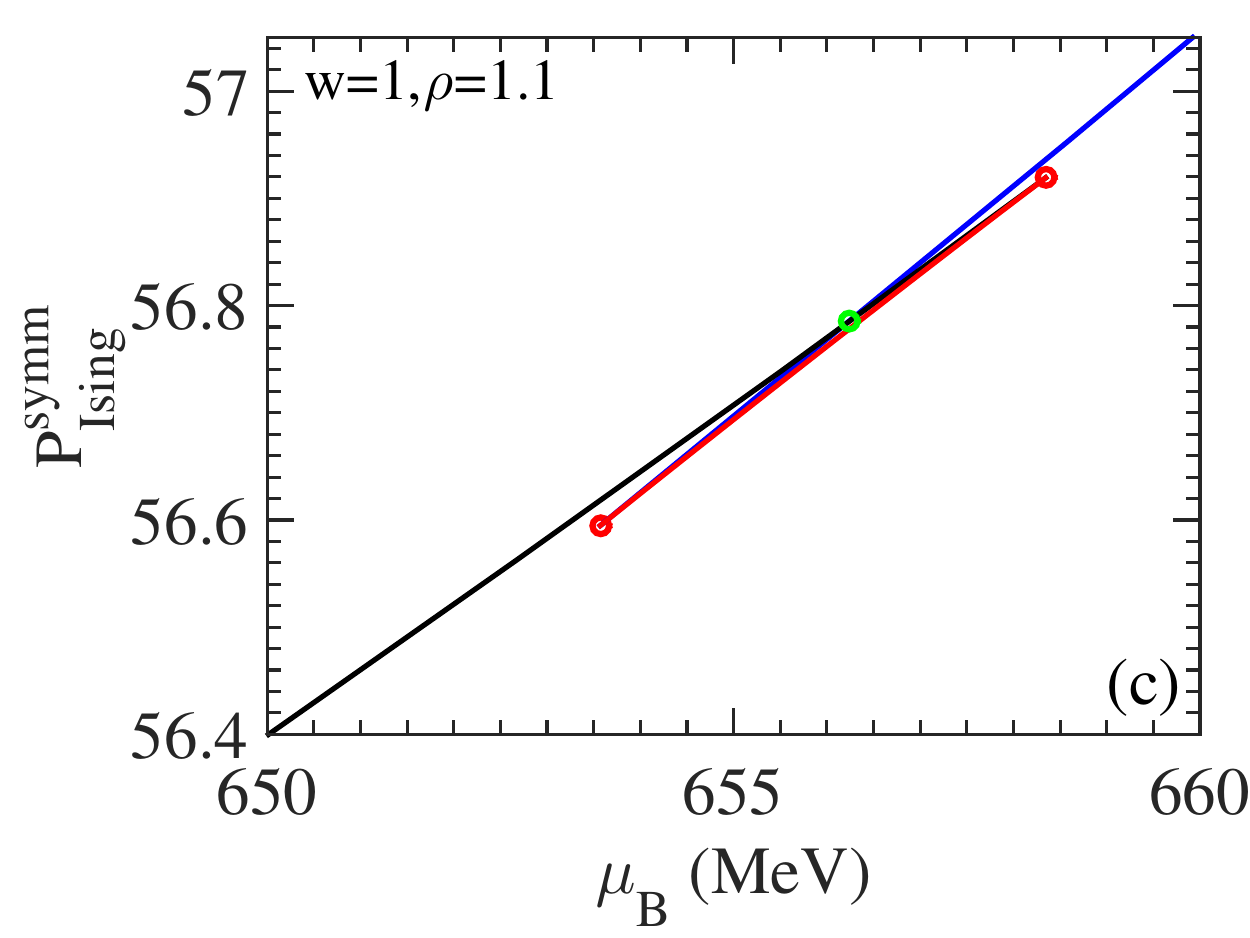}	
	\includegraphics[width=0.32\textwidth]{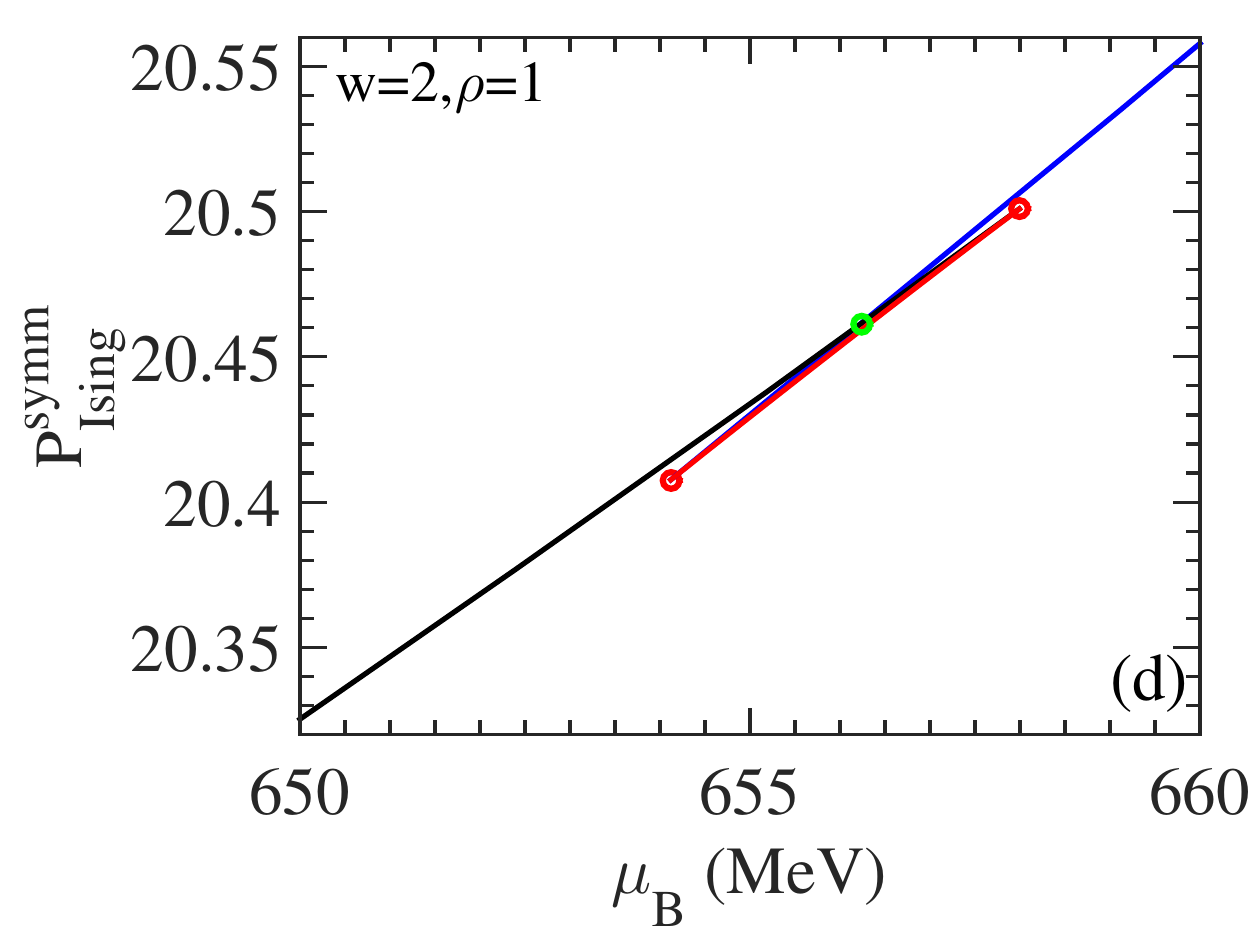}
	\includegraphics[width=0.32\textwidth]{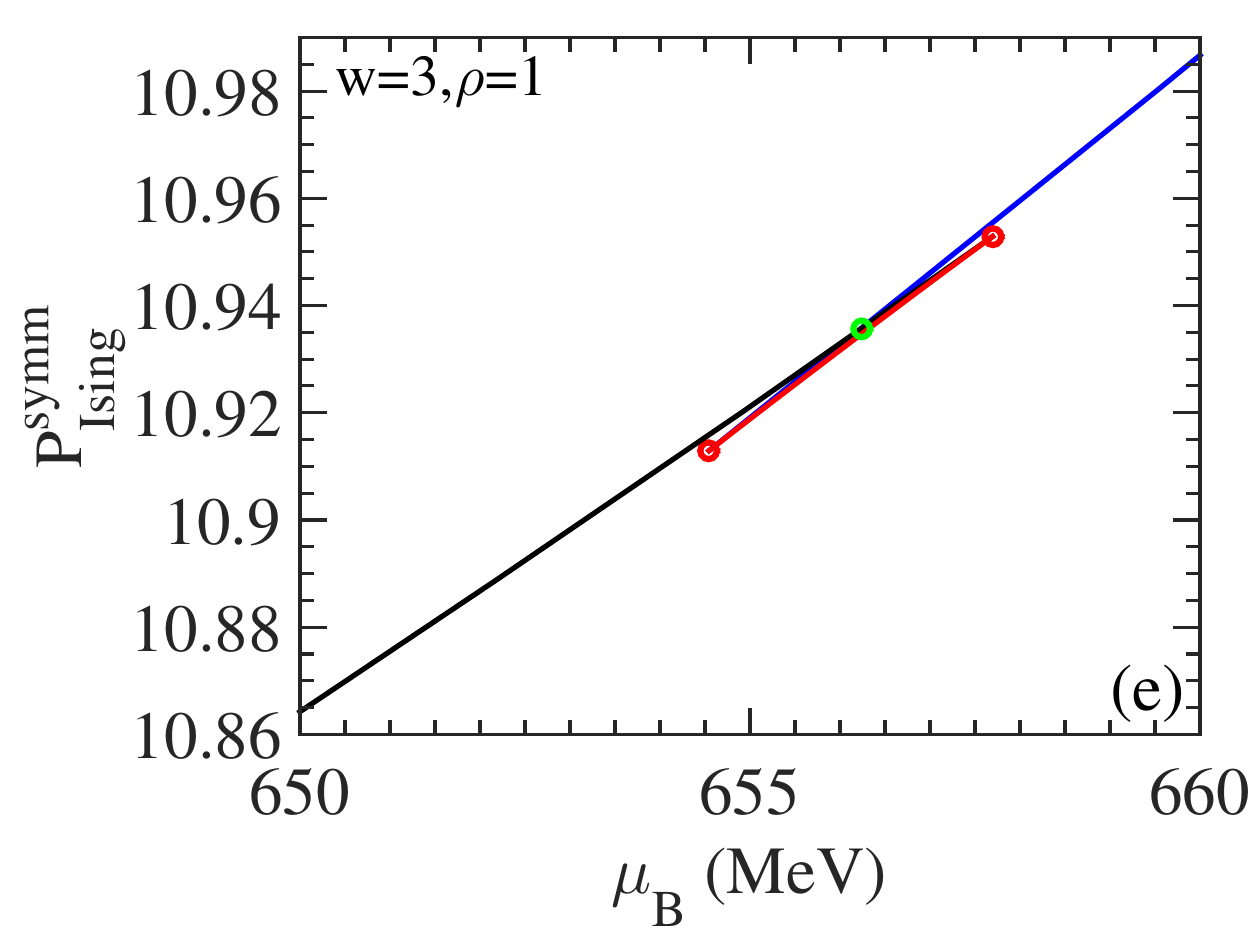}
	\caption{\label{Fig. 6}(Color online). Symmetrized Ising model pressure as a function of the net-baryon chemical potential for $\alpha_2=\alpha_1-6^\circ$ at several values of the scaling parameters. The open green and red circles indicate the location of the coexistence and spinodal points, respectively.}
\end{figure*}

In Figs.~5(a), (b) and (c), $w$ is fixed at $1$, while $\rho$ takes values of $1$, $2$ and $3$, respectively. In Figs.~5(d) and (e), $\rho$ is fixed at $2$, while $w$ takes values of $2$ and $3$, respectively. Evidently, the value of $\mu_B$ at the coexistence point is independent of both $w$ and $\rho$. For the spinodal points, when both $w$ and $\rho$ take small values, for instance, in Fig.~5(a) with $w=1$ and $\rho=1$, only a single spinodal point emerges; the intersection of the second and third solutions is not captured. This observation is consistent with Fig.~3(a): the isotherm at $0.95T_C$ yields only one crossing point with the cyan dashed line. 

As $\rho$ increases, as shown in Fig.~5(b) and (c), two spinodal points emerge. Their separation in $\mu_B$ decreases, and they approach each other as well as the coexistence point. Meanwhile, $P_{\mathrm{Ising}}^{\mathrm{symm}}$ rises with increasing $\rho$. 

With $\rho$ fixed at $2$, $w$ takes values of $1$, $2$ and $3$ as illustrated in Figs.~5(b) and (d) and (e), respectively, two spinodal points are present in all cases. The $\mu_B$ separation between these points likewise decreases with increasing $w$, and the spinodal points move toward one another as well as toward the coexistence point. Meanwhile, $P_{\mathrm{Ising}}^{\mathrm{symm}}$ decreases as $w$ grows.

The $\mu_B$ dependence of $P_{\mathrm{Ising}}^{\mathrm{symm}}$ along the isotherm $T=0.95T_C$ after mapping to the QCD phase diagram with $\alpha_2=\alpha_1-6^\circ$ is shown in Fig.~6. The $\mu_B$ value ranges from $650$ to $660$ MeV. The three pressure solutions are also plotted in distinct colors as done in Fig.~5. 
In Figs.~6(a), (b) and (c), $w$ is fixed at $1$, while $\rho$ equals $0.9$, $1$ and $1.1$, respectively. In Figs.~6(d) and (e), $\rho$ is fixed at $1$, while $w$ takes values of $2$ and $3$, respectively.

Firstly, $\mu_B$ at the coexistence point is independent of both $w$ and $\rho$. Two spinodal points can be identified in each subfigure. The first solution (plotted in blue) intersects the third solution (plotted in red) at the spinodal point with a smaller $\mu$, while the second solution (plotted in black) intersects the third solution at the spinodal point with a larger $\mu$. This behavior is reversed for $\alpha_2=\alpha_1+90^\circ$. Secondly, their separation in $\mu_B$ decreases with increasing $\rho$ or $w$. This qualitative trend is consistent with that observed in Fig.~5. Thirdly, $P_{\mathrm{Ising}}^{\mathrm{symm}}$ decreases with increasing $\rho$ as well as increasing $w$. This differs from the case $\alpha_2=\alpha_1+90^\circ$, where $P_{\mathrm{Ising}}^{\mathrm{symm}}$ rises with increasing $\rho$ but decreases with increasing $w$. What is more, the values of $P_{\mathrm{Ising}}^{\mathrm{symm}}$ obtained from the three solutions are very close to one another at the same $\mu_B$ and their slopes are also comparable.

\begin{figure*}[hbt]
	\centering
	\includegraphics[width=0.32\textwidth]{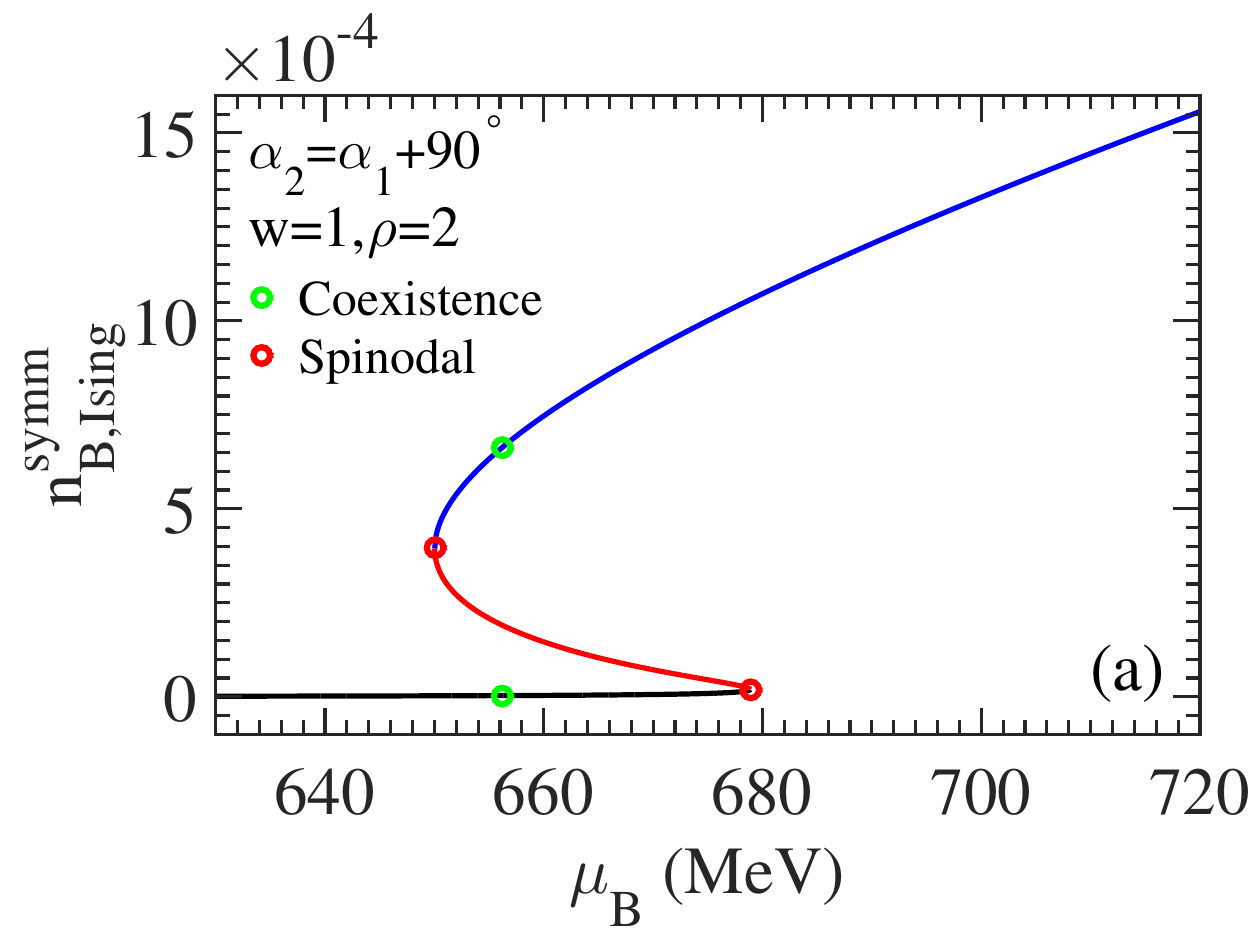}
	\includegraphics[width=0.32\textwidth]{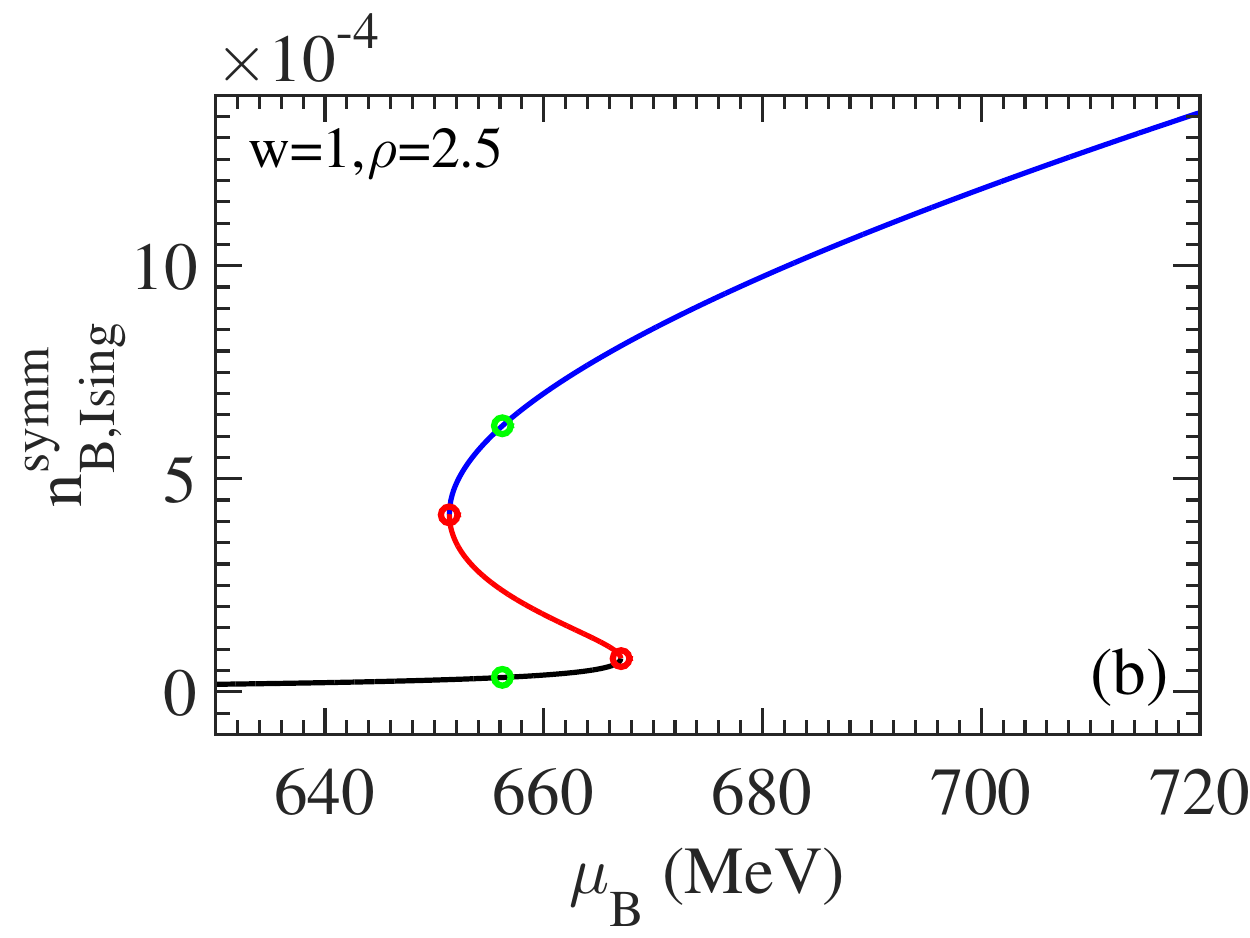}
	\includegraphics[width=0.32\textwidth]{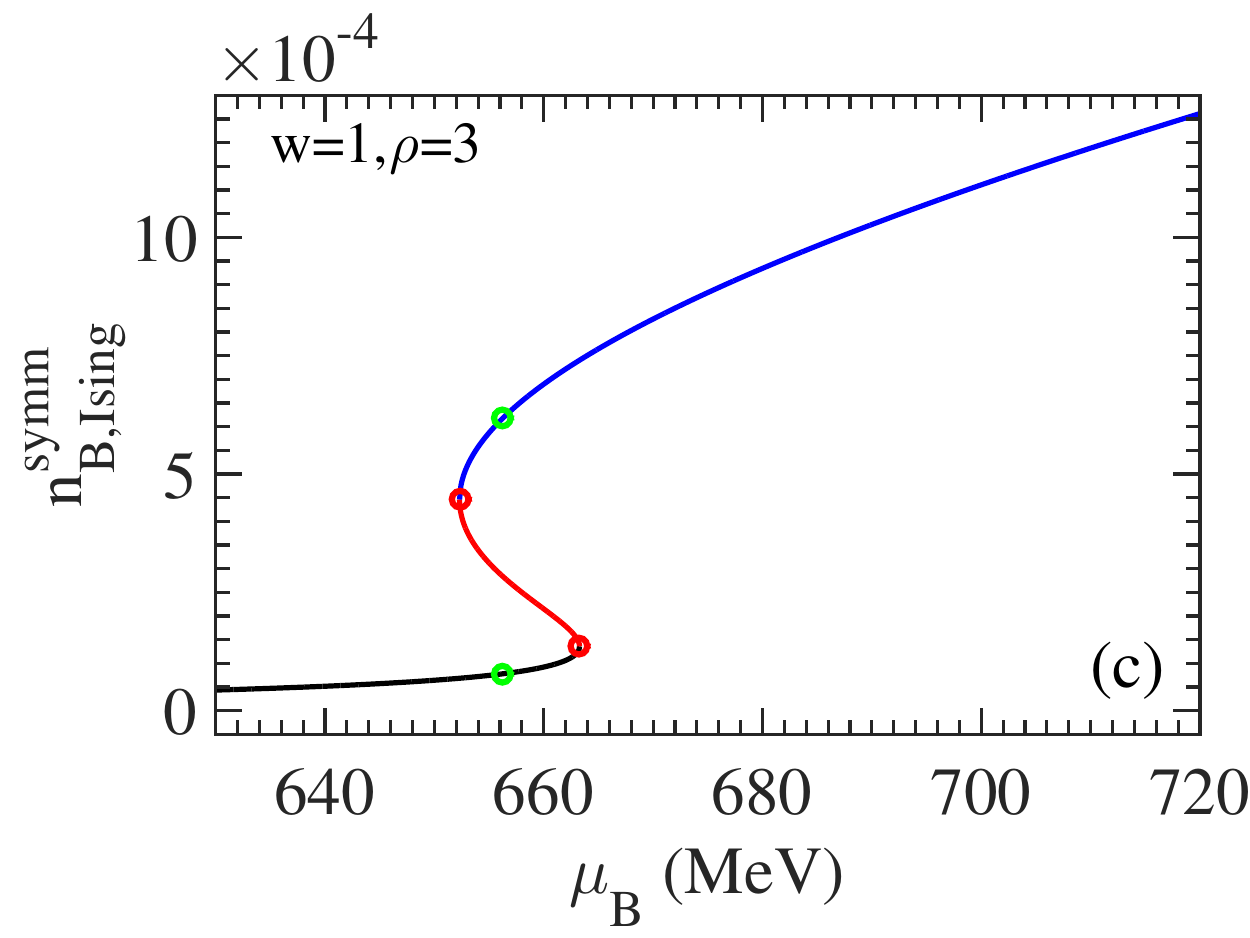}
	\includegraphics[width=0.32\textwidth]{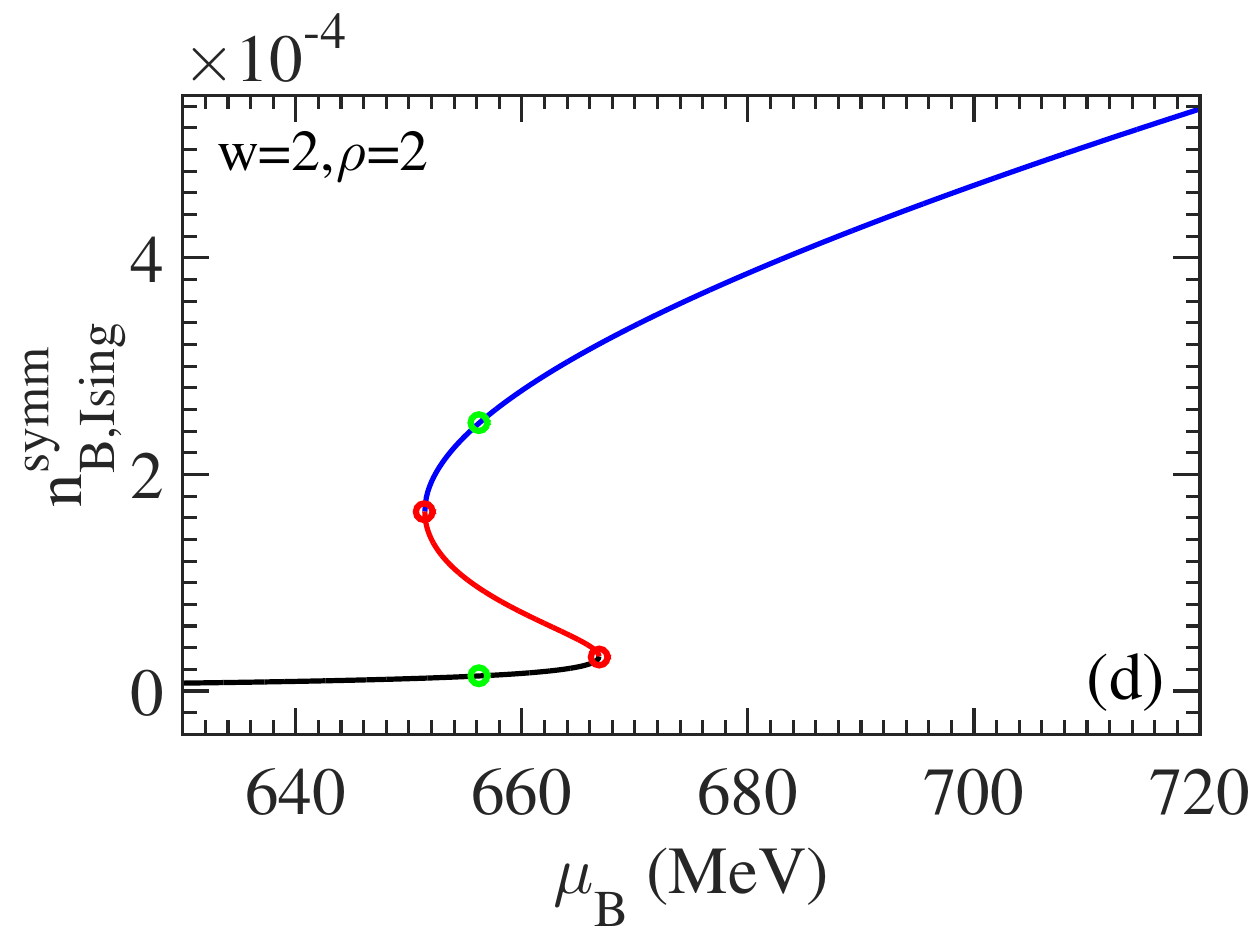}
	\includegraphics[width=0.32\textwidth]{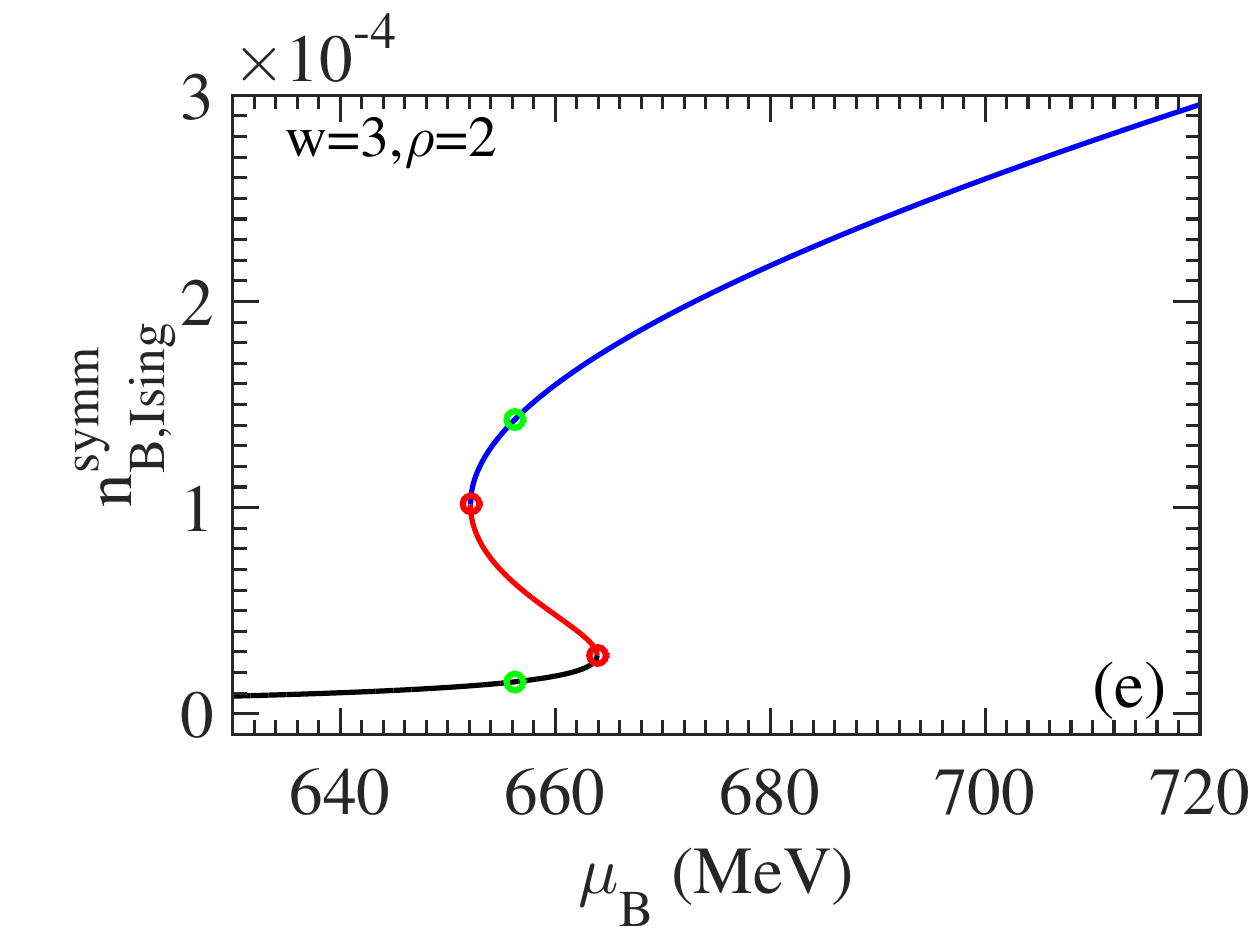}
	
	\caption{\label{Fig. 7}(Color online). Symmetrized Ising model density as a function of the net-baryon chemical potential for $\alpha_2=\alpha_1+90^\circ$ at different scaling parameters. The open green and red circles indicate the location of the coexistence and spinodal points, respectively. }
\end{figure*}

\begin{figure*}[hbt]
	\centering
	\includegraphics[width=0.32\textwidth]{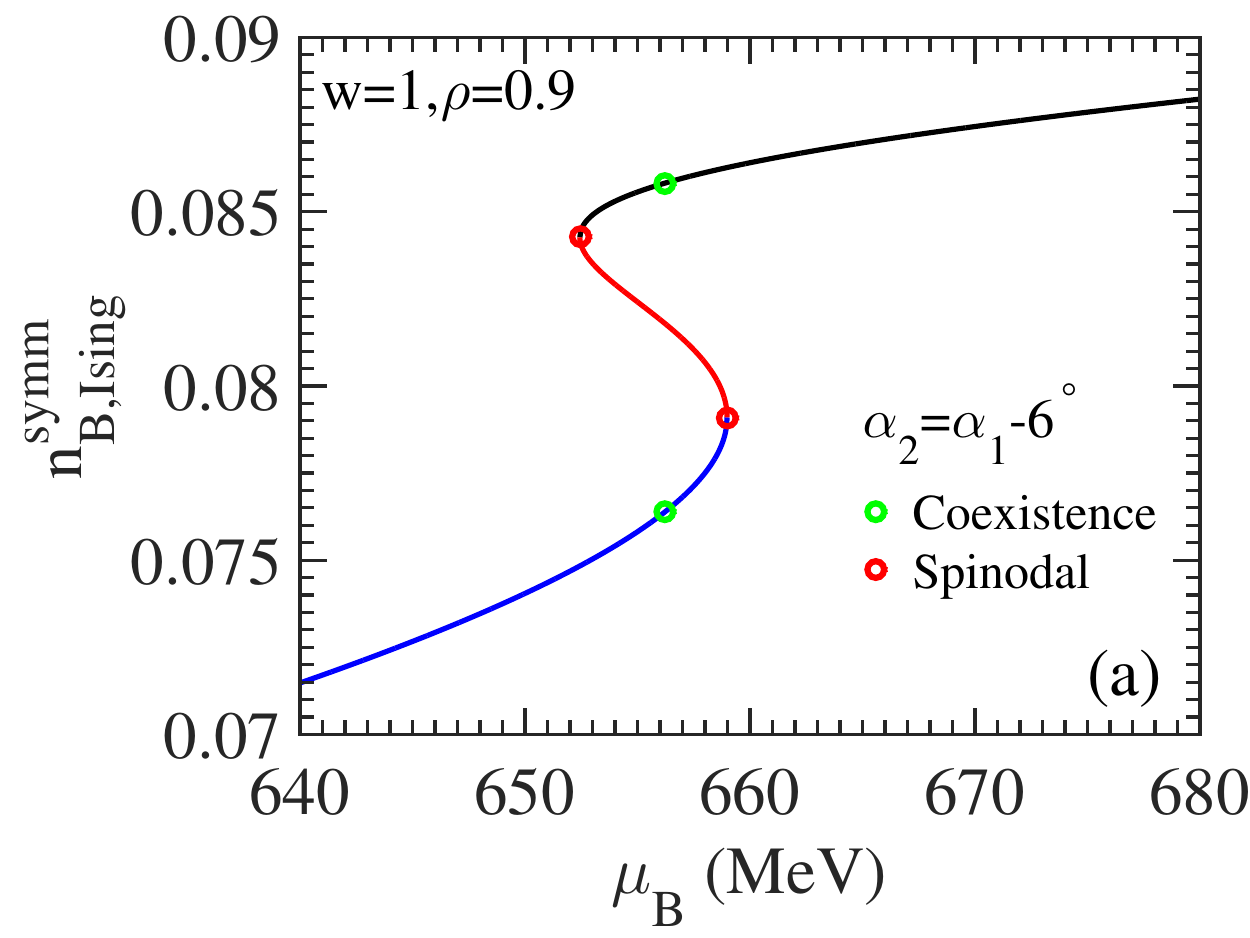}
	\includegraphics[width=0.32\textwidth]{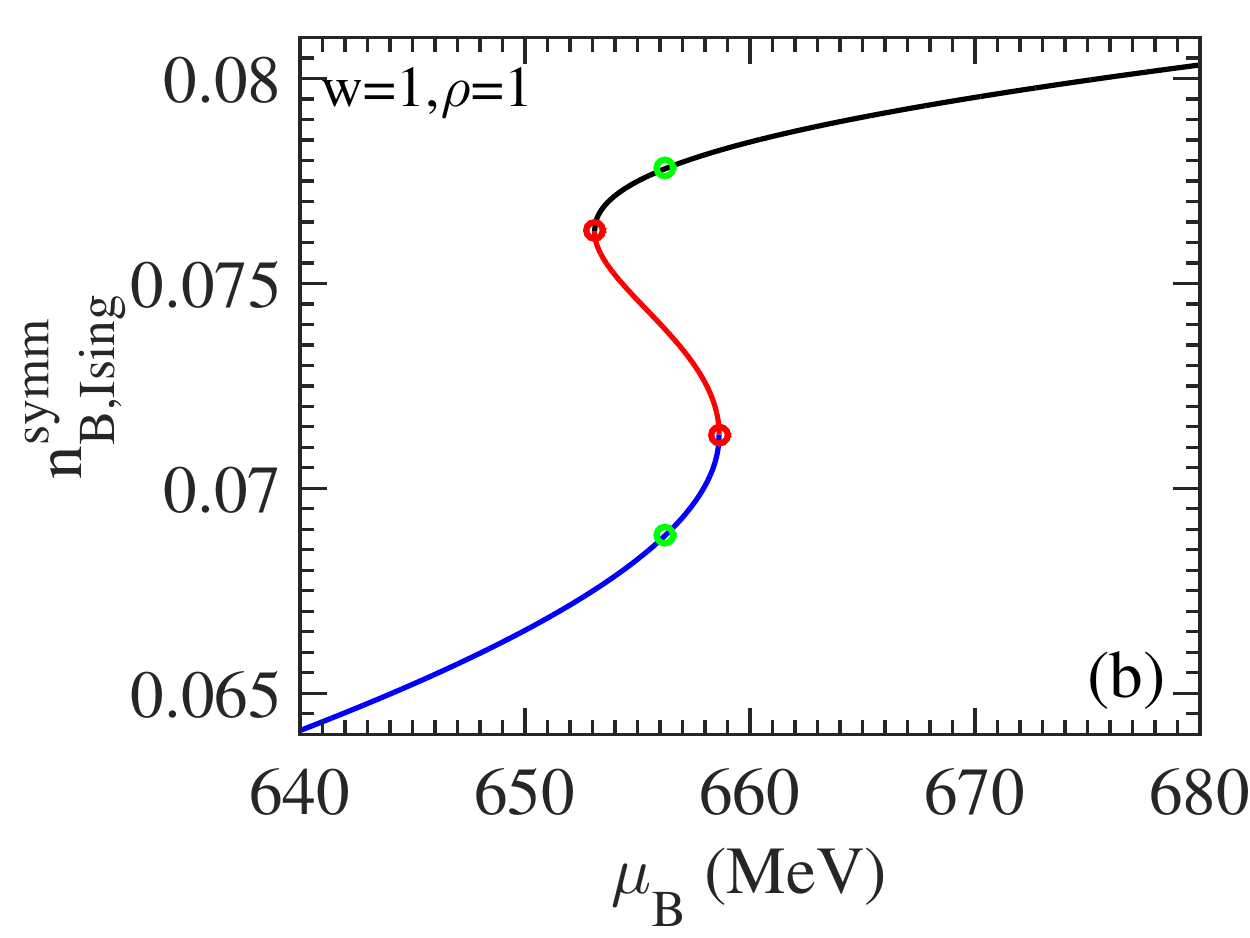}
	\includegraphics[width=0.32\textwidth]{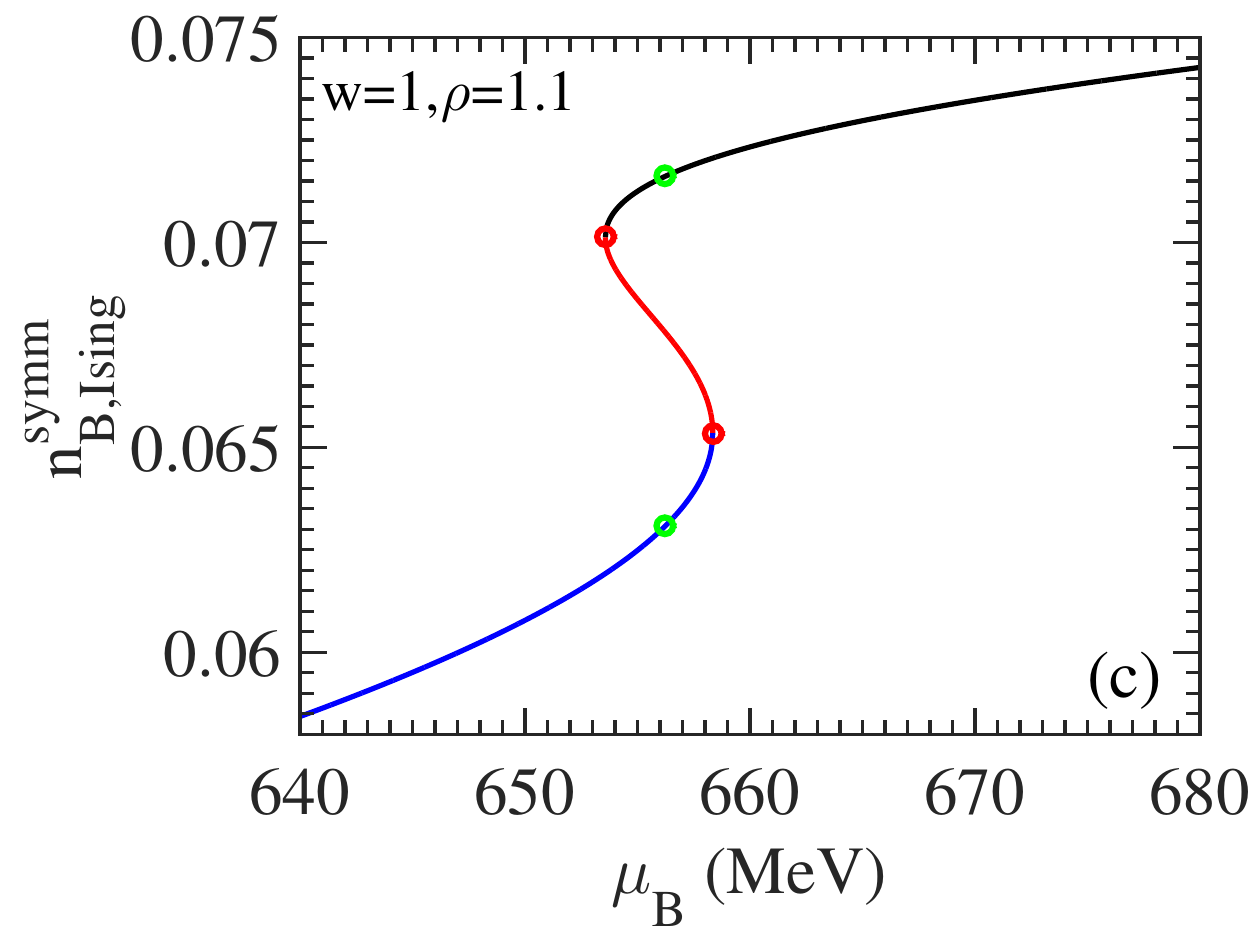}
	\includegraphics[width=0.32\textwidth]{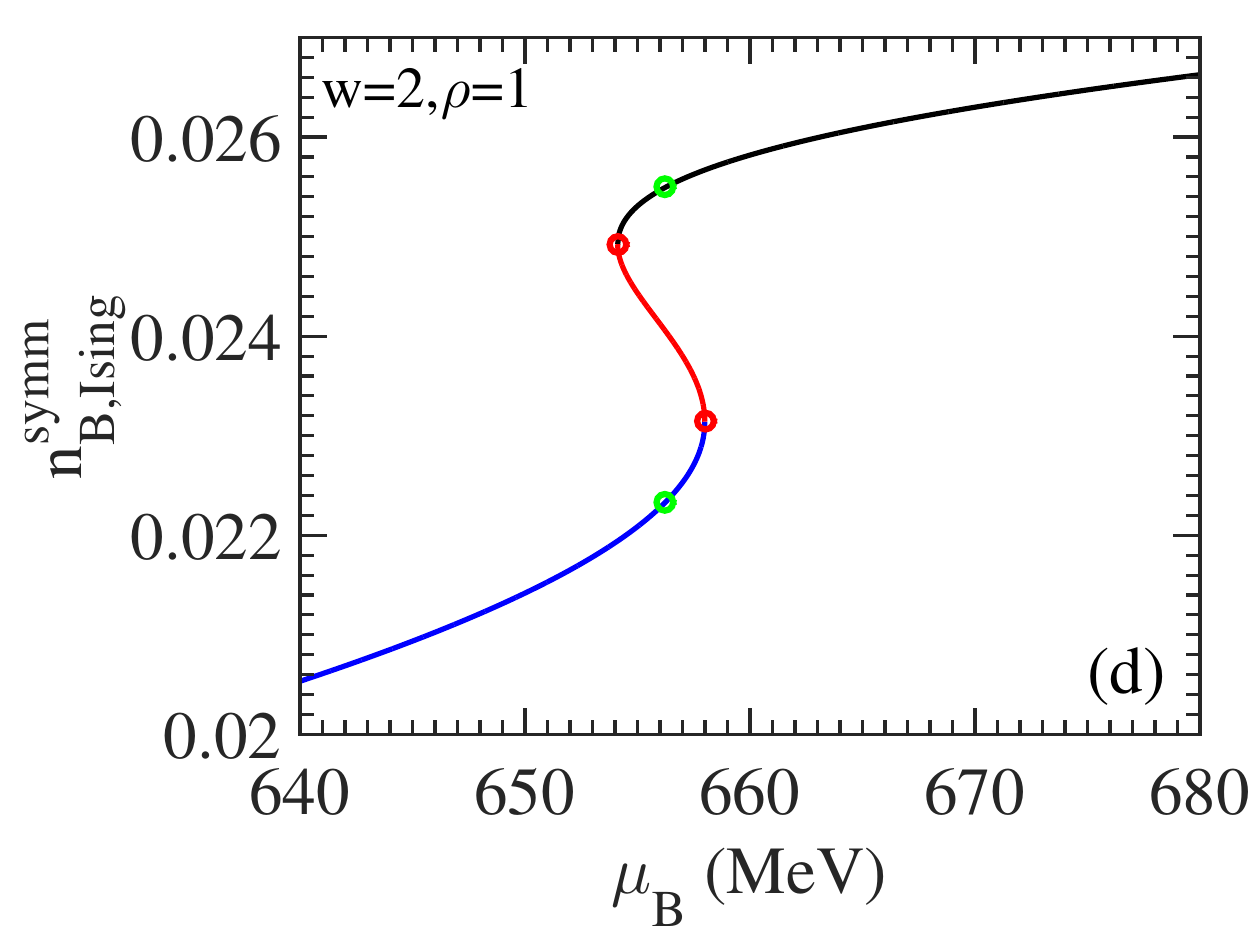}
	\includegraphics[width=0.32\textwidth]{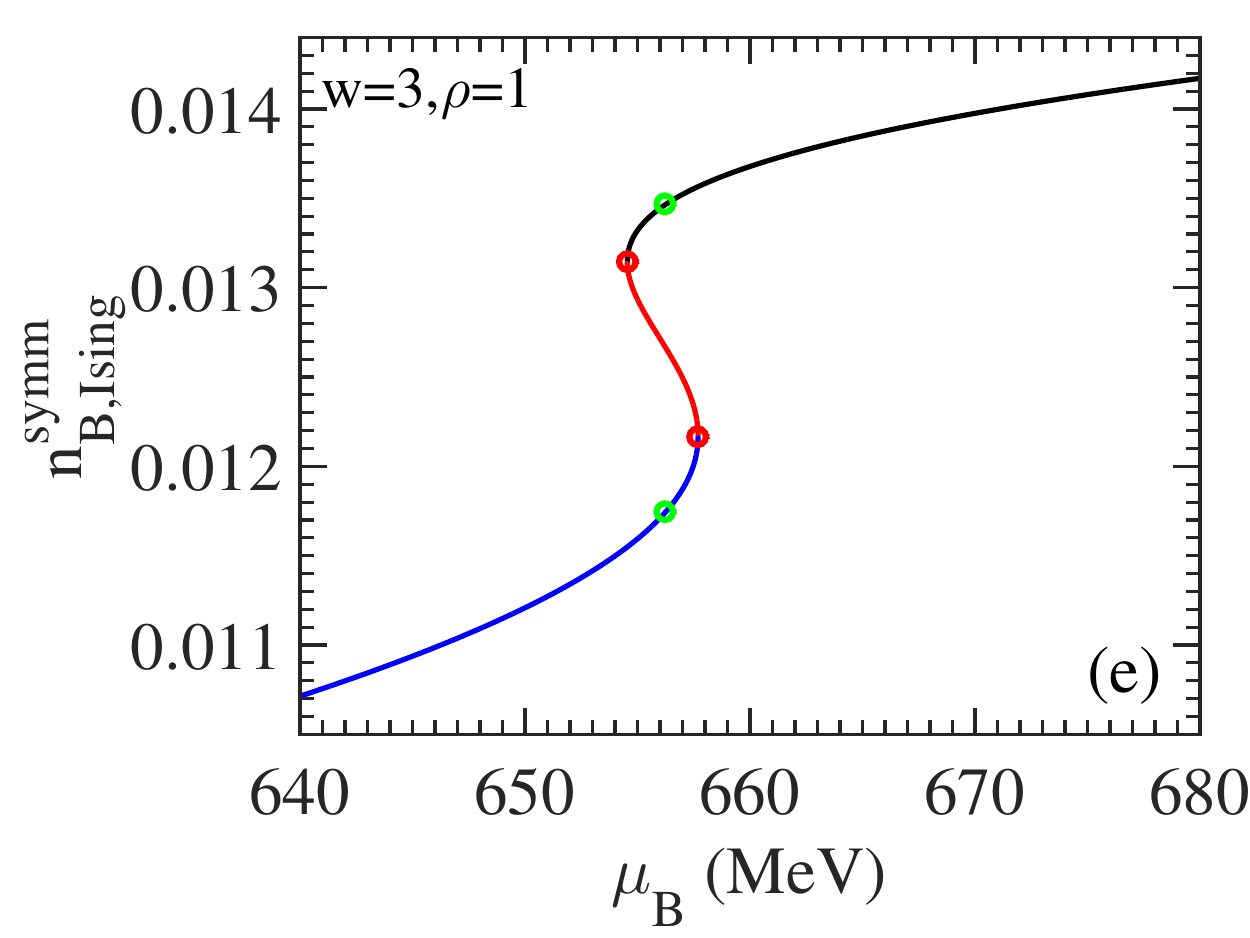}
	\caption{\label{Fig. 8}(Color online). Symmetrized Ising model density as a function of the net-baryon chemical potential for $\alpha_2=\alpha_1-6^\circ$ at different scaling parameters. The open green and red circles indicate the location of the coexistence and spinodal points, respectively.}
\end{figure*}

To directly visualize the metastable and unstable regions, the $\mu_B$ dependence of symmetrized density from the Ising model ($n_{B,\mathrm{Ising}}^{\mathrm{symm}}$) after mapping to the QCD phase diagram along the isotherm $0.95T_C$ for $\alpha_2=\alpha_1+90^\circ$ is shown in Fig.~7.

In Figs.~7(a), (b) and (c), $w$ is fixed at $1$, while $\rho$ takes values of $2$, $2.5$ and $3$, respectively. In Figs.~7(d) and (e), $\rho$ is fixed at $2$, while $w$ takes values of $2$ and $3$, respectively. As $\rho$ or $w$ increases, both the metastable region (the $\mu_B$ separations between the open green circle and the adjacent open red circles along the isotherm) and unstable region (the $\mu_B$ separations between the open red circles) become narrower. As $\rho$ increases, $n_{B,\mathrm{Ising}}^{\mathrm{symm}}$ exhibits only a small change. For instance, $n_{B,\mathrm{Ising}}^{\mathrm{symm}}$ at the upper green circle exhibits only a small change when moving from Fig.~7(a) to (c). As $w$ increases, $n_{B,\mathrm{Ising}}^{\mathrm{symm}}$ decreases. Similarly, $n_{B,\mathrm{Ising}}^{\mathrm{symm}}$ at the upper green circle decreases across Fig.~7(a), (d) and (e). 

$n_{B,\mathrm{Ising}}^{\mathrm{symm}}$ for $\alpha_2=\alpha_1-6^\circ$ is shown in Fig.~8. In Figs.~8(a), (b) and (c), $w$ is fixed at $1$, while $\rho$ takes values of $0.9$, $1.0$ and $1.1$, respectively. In Figs.~8(d) and 8(e), $\rho$ is fixed at $1$, while $w$ takes values of $2$ and $3$, respectivly. The effect of $\rho$ and $w$ on the metastable region and unstable spinodal region remains the same as that for $\alpha_2=\alpha_1+90^\circ$. As $\rho$ or $w$ increases, both the metastable region and unstable region become narrower. As $\rho$ or $w$ increases, $n_{B,\mathrm{Ising}}^{\mathrm{symm}}$ decreases. For instance, $n_{B,\mathrm{Ising}}^{\mathrm{symm}}$ at the upper green circle decreases when moving from Fig.~8(a) to (c) or across Fig.~8(b), (d) and (e).   

In summary, qualitatively, for the two mapping schemes of the mean-field Ising model onto QCD, increasing the scaling parameters $\rho$ or $w$ does not shift the positions of the coexistence points but narrows the metastable and unstable regions. Quantitatively, increasing $w$ reduces both $P_{\mathrm{Ising}}^{\mathrm{symm}}$ and $n_{B,\mathrm{Ising}}^{\mathrm{symm}}$. For $\alpha_2=\alpha_1-6^\circ$, increasing $\rho$ also reduces both $P_{\mathrm{Ising}}^{\mathrm{symm}}$ and $n_{B,\mathrm{Ising}}^{\mathrm{symm}}$, whereas for $\alpha_2=\alpha_1+90^\circ$, increasing $\rho$ raises $P_{\mathrm{Ising}}^{\mathrm{symm}}$ and has only a small effect on $n_{B,\mathrm{Ising}}^{\mathrm{symm}}$.

\section{Summary}

In this work, we study the effects of scaling parameters on the mapping of the mean-field Ising model onto QCD for two mapping schemes: the conventional one with $\alpha_2=\alpha_1+90^{\circ}$, and the one valid in the small quark mass limit with $\alpha_2=\alpha_1-6^{\circ}$. We discuss how these parameters affect the metastable and unstable regions as well as the symmetrized pressure and density along the isotherm $T=0.95T_C$ on the QCD $T-\mu_B$ phase diagram.

For both the conventional mapping scheme and the scheme valid in the small-quark-mass-limit, the $\mu_B$ values of the coexistence points are independent of the scaling parameters $\rho$ and $w$. When $\rho$ and $w$ are sufficiently large, two spinodal points appear on the isotherm, one on each side of the coexistence line. As $\rho$ or $w$ increases, the spinodal points move toward each other and toward the coexistence line, which results in narrower metastable and unstable regions along the $\mu_B$ direction.

Near the critical point, for identical scaling parameters $\rho$ and $w$, the spinodal points along the isotherm obtained with the small-quark-mass-limit mapping scheme lie closer to one another and to the coexistence point than those from the conventional mapping scheme. 

In addition, the symmetrized pressure and density vary with the scaling parameters. For both mapping schemes, increasing $w$ decreases $P_{\mathrm{Ising}}^{\mathrm{symm}}$ and $n_{B,\mathrm{Ising}}^{\mathrm{symm}}$. For $\alpha_2=\alpha_1-6^\circ$, raising $\rho$ also reduces both quantities. In contrast, for $\alpha_2=\alpha_1+90^\circ$, increasing $\rho$ raises $P_{\mathrm{Ising}}^{\mathrm{symm}}$ and has only a small effect on $n_{B,\mathrm{Ising}}^{\mathrm{symm}}$.

\vskip 0.5cm
This work is supported by Research Start-up Fund Project of Chengdu University (X2110) and the National Supercomputing Center in Chengdu-Chengdu University Branch.

\ed